\documentclass[dvipsnames,twocolumn]{aastex631}
\usepackage{placeins}
\usepackage{graphicx}
\usepackage{amsmath}
\usepackage{txfonts}
\usepackage{color}

\graphicspath{{Figures/}}

\newcommand{\mdisk}[0]{\ensuremath{M_{\rm disk}}}

\newcommand{\mdust}[0]{\ensuremath{M_{\rm dust}}}
\newcommand{\rgas}[0]{\ensuremath{R_{\rm CO,\ 90\%}}}
\newcommand{\rc}[0]{\ensuremath{R_{\rm c}}}
\newcommand{\macc}[0]{\ensuremath{\dot{M}_{\rm acc}}}
\newcommand{\mstar}[0]{\ensuremath{M_{*}}}
\newcommand{\msun}[0]{\ensuremath{\mathrm{M}_{\odot}}}

\newcommand{\ro}[0]{\ensuremath{R_{\rm c,0}}}
\newcommand{\mo}[0]{\ensuremath{M_0}}
\newcommand{\tacc}[0]{\ensuremath{t_{\rm acc,0}}}
\newcommand{\aDW}[0]{\ensuremath{\alpha_{DW}}}
\newcommand{\aSS}[0]{\ensuremath{\alpha_{SS}}}
\newcommand{\rout}[0]{\ensuremath{R_{\rm out}}}

\begin{document}

\title{Effect of MHD wind-driven disk evolution on the observed sizes of protoplanetary disks}

\correspondingauthor{Leon Trapman}
\email{ltrapman@wisc.edu}
\author[0000-0002-8623-9703]{Leon Trapman}
\affiliation{Department of Astronomy, University of Wisconsin-Madison, 
475 N Charter St, Madison, WI 53706}

\author[0000-0002-1103-3225]{Beno\^it Tabone}
\affiliation{Leiden Observatory, Leiden University, 2300 RA Leiden, the Netherlands}

\author[0000-0003-4853-5736]{Giovanni Rosotti}
\affiliation{Leiden Observatory, Leiden University, 2300 RA Leiden, the Netherlands}
\affiliation{School of Physics and Astronomy, University of Leicester, Leicester LE1 7RH, UK}

\author[0000-0002-0661-7517]{Ke Zhang}
\affiliation{Department of Astronomy, University of Wisconsin-Madison, 
475 N Charter St, Madison, WI 53706}

\begin{abstract}
It is still unclear whether the evolution of protoplanetary disks, a key ingredient in the theory of planet formation, is driven by viscous turbulence or magnetic disk winds.
As viscously evolving disks expand outward over time, the evolution of disk sizes is a discriminant test for studying disk evolution. 
However, it is unclear how the observed disk size changes over time if disk evolution is driven by magnetic disk winds.
Combining the thermochemical code DALI with the analytical wind-driven disk evolution model presented in \cite{Tabone2021a}, we study the time evolution of the observed gas outer radius as measured from CO rotational emission (\rgas). 
The evolution of \rgas\ is driven by the evolution of the disk mass, as the physical radius stays constant over time.
For a constant \aDW, an extension of the $\alpha-$Shakura-Sunyaev parameter to wind-driven accretion, \rgas\ decreases linearly with time. Its initial size is set by the disk mass and the characteristic radius \ro, but only \ro\ affects the evolution of \rgas, with a larger \ro\ resulting in a steeper decrease of \rgas.
For a time-dependent \aDW, \rgas\ stays approximately constant during most of the disk lifetime until \rgas\ rapidly shrinks as the disk dissipates. 
The constant \aDW-models are able to reproduce the observed gas disk sizes in the $\sim1-3$ Lupus and $\sim5-11$ Myr old Upper Sco star-forming regions. However, they likely overpredict the gas disk size of younger $(\lessapprox0.7\ \mathrm{Myr})$ disks.
\end{abstract}

%
\section{Introduction}
\label{sec: introduction}

Planets are formed in gas-rich disks around young stars. The processes through which these planets are formed are still not fully understood (see, e.g. \citealt{Benz2014,MorbidelliRaymond2016,Morton2016,Mordasini2018}). A key ingredient of planet formation models is the behavior of the gas in these planet-forming disks. 
Being the dominant mass component in disks, the gas plays a critical role in nearly all processes that lead to planet formation. The total gas mass represents the reservoir available for forming giant planets. The gas density and local gas-to-dust mass ratio regulate the dynamics of dust grains and larger solid bodies in the disk, setting the rate at which the dust grows, settles towards the midplane and drift inwards toward the star (e.g. \citealt{Wada2008,Birnstiel2010,Birnstiel2012,Krijt2015}). 
Furthermore, these same processes also affect how quickly planets can accrete solids in the pebble accretion view (see, e.g. \citealt{Bitsch2015,OrmelLiu2018}).
Understanding the evolution of the gas is therefore crucial for increasing our understanding of planet formation.

The evolution of the gas is set by the transport of angular momentum that drives the stellar mass accretion flow. Two mechanisms have been proposed to drive this process: turbulent viscosity and magneto-hydrodynamical (MHD) disk winds. 
Of these two, viscosity is commonly thought to be the dominant process (see, e.g. \citealt{Armitage2019}). In the viscous evolution framework the disk evolves as a result of turbulence acting as an effective viscosity that redistributes the angular momentum (see e.g. \citealt{LyndenBellPringle1974,ShakuraSunyaev1973,Pringle1981}). Most of the angular momentum is transported outward radially with a small fraction of the mass, which results in the remaining mass moving inward, where it is accreted onto the star. This behavior is often referred to as \emph{viscous spreading}. Viscous evolution has been successfully used to explain the observed correlation between stellar mass accretion rate and disk mass (see, e.g. \citealt{Manara2016b,Rosotti2017,Lodato2017}).
The high levels of turbulence required to drive the stellar accretion have been explained by the magneto-rotational instability (MRI; see, e.g. \citealt{BalbusHawley1991,BalbusHawley1998}). 
However, numerical simulations including detailed microphysics show that MRI is quenched in large regions of the disk due to non-ideal MHD effects (so-called ``dead-zones'', see, e.g. \citealt{Gammie1996, BaiStone2011, Bai2011}).
Furthermore, recent observations show that turbulence in outer parts of protoplanetary disks in minimal, making it unclear whether disks are turbulent enough to drive the observed accretion rate (see, e.g. \citealt{Flaherty2015,Flaherty2018,Flaherty2020}).

Magneto-hydrodynamical (MHD) disk winds appear to be a compelling alternative to drive stellar mass accretion.
A magnetic field with a net flux present in the disk could launch material from the surface disk along field lines, resulting in a rotating disk wind (see e.g. \citealt{BlandfordPayne1982,Turner2014,Lesur2020}). The launched material extracts angular momentum from the disk, causing the remaining material to spiral inward towards the star. 
Simulations and (semi-)analytical models show that these winds are able to extract enough angular momentum to drive stellar accretion (see, e.g. \citealt{Ferreira2006,BaiStone2013,Bethune2017,ZhuStone2017,Lesur2020}). However, observations of these rotating disk winds are sparse and it is unclear to what degree they contribute to disk evolution on average (see, e.g., \citealt{Tabone2017,deValon2020}). 
Note that, in contrast to viscous evolution, MHD wind-driven evolution does not require the disk to be turbulent. Whether disk evolution is driven by viscosity or disk winds therefore has a profound effect on planet formation.

Direct detections of the processes that drive disk evolution have proven difficult (see, e.g. \citealt{Flaherty2015,Flaherty2018,Flaherty2020,Tabone2017,Najita2021}), leading us to instead study the effect of disk evolution on the global properties of the disk (see, e.g. \citealt{Manara2016b,Rosotti2017,Lodato2017}). In a recent study, \cite{Tabone2021b} show that MHD disk winds can account for the fast disk dispersal (following the pioneering work of \citealt{Armitage2013}), and the correlation between stellar mass accretion rates and disk masses. In particular, observations of the evolution of the size of the gas disk can provide an discriminant diagnostic. Both evolutionary scenarios, viscous and wind-driven, make different predictions on how the gas disk size changes over time. For a viscously evolving disk the aforementioned viscous spreading will cause the outer part of the disk to expand outward. In other words, the disk will grow in size over time (but see \citealt{YangBai2021} for the possibility of wind-driven disks growing over time). 
For a MHD wind-driven disk this is not the case. 
Disk accretion is driven by the angular momentum extracted vertically by the wind, 
meaning there is no need for viscous spreading. Instead, a wind-driven disk is expected to stay the same size or even shrink over time.
To distinguish between these two scenarios, we need to measure disk sizes for a large number of disks that spread a wide range of disk ages.

Recently, these observations have started to become available, thanks in no small part to large disk surveys carried out with the Atacama Large Millimeter/sub-millimeter Array (ALMA).  
Using ALMA, gas disk sizes, defined as the radius that encloses 90\% of the $^{12}$CO $J\,=\,2\,-\,1$ emission, have been uniformly measured for 35 disks in the 1-3 Myr old Lupus star-forming region \citep{ansdell2018,Sanchis2021}. Similarly, \cite{barenfeld2017} measured gas disk sizes for 9 disks in older 5-11 Myr Upper Sco star-forming region.

Analysis of these observed gas disk sizes have so far focused predominantly on viscous evolution. \cite{NajitaBergin2018} collected and analyzed a sample of disk sizes from the literature. They showed that older Class II protoplanetary disks are overall larger than younger Class I embedded disks, which is consistent with the prediction for viscous evolution. However, the variety of different observational tracers and definitions for observed gas disk size in their sample makes it difficult to quantify their results.  
To address this issue, \cite{Trapman2020} used physical-chemical models to perform a quantitative analysis of the link between \rgas\ and $\alpha_{\rm visc}$, the dimensionless $\alpha$-parameter often used to parameterize viscosity in disks (see \citealt{ShakuraSunyaev1973}).
They show that the observed \rgas\ of disks in Lupus are consistent with having evolved viscously with a low viscosity $(\alpha_{\rm visc} = 10^{-4} - 10^{-3})$. However, they also show that the observed \rgas\ of disks in older Upper Sco, which are on average smaller than disks in Lupus, cannot be explained by viscous evolution alone. Some other process, such as external photo-evaporation, is required to explain the combined set of \rgas\ of disks in Lupus and Upper Sco with viscous evolution.

A similar analysis of \rgas\ from the perspective of MHD disk wind-driven evolution is currently lacking. While the disk wind is not expected to have an effect on the physical size of the disk, it does affect the evolution of the disk mass, gas surface density, temperature structure and chemistry, which all affect the emission from which \rgas is measured in observations. It is therefore not immediately clear how the observed disk size changes over time for a MHD wind-driven disk.

In this work we combine the analytical model for MHD wind-driven disk evolution recently developed by \cite{Tabone2021a}, with the thermochemical code \texttt{DALI} \citep{Bruderer2012,Bruderer2013} to investigate how \rgas\ evolves in this scenario. 
This manuscript is structured as follows: In Section \ref{sec: Model setup} we outline the analytical model and explain how we link it to \texttt{DALI}. In Section \ref{sec: results} we examine the evolution of \rgas, first using a toy model to discuss the differences between viscous and wind-driven disk evolution and next using \texttt{DALI} models to include the effects of CO chemistry and excitation. We compare our models to the observations in Lupus and Upper Sco in Section \ref{sec: discussion}, where we also discuss the assumptions and caveats in our modeling approach. Finally, we summarize our results in Section \ref{sec: conclusions}.

\section{Model setup}
\label{sec: Model setup}

\subsection{Disk wind $\alpha-$formalism}
\label{sec: wind-driven disk evolution}

In this work we use the analytical model for disk evolution driven by MHD disk winds presented by \cite{Tabone2021a}. Here we briefly summarize the assumptions they have made to derive their model and show the results relevant for this work. For a more detailed description we refer the reader to \cite{Tabone2021a}. Their model is a 
1D global evolutionary model for MHD wind-driven disk evolution similar to the model for viscous evolution presented by \cite{ShakuraSunyaev1973} and \cite{LyndenBellPringle1974}. As with the viscous-$\alpha$ model, it makes minimal assumptions on any physical processes that drive the transport and removal of angular momentum, but instead provides a framework to quantify their average efficiency.

The MHD disk wind has two effects on the evolution of the surface density of the disk. The wind extracts angular momentum from the disk, which causes material to move inward and drives the mass accretion onto the star. In order for the angular momentum to be extracted some material must also be carried away. The second effect is therefore the mass-loss rate caused by the wind. Combining these two effects with the turbulent transport of angular momentum known from viscous evolution, the time evolution of the surface density $\Sigma$ can be written as (Eq. 10 in \citealt{Tabone2021a}, see their paper for its derivation)

\begin{align}
\label{eq: master equation}
\partial_t\Sigma &= \frac{3}{r}\partial_r\left[\frac{1}{r\Omega}\partial_r\left(r^2\aSS\Sigma c_s^2\right)\right] \\
                 &+ \frac{3}{2r}\partial_r\left[ \frac{\aDW\Sigma c_s^2}{\Omega}\right] \\
                 &- \frac{3\aDW \Sigma c_s^2}{4\left( \lambda - 1\right)r^2\Omega}.
\end{align}

Here the first term on the right hand side describes the radial redistribution of angular momentum by turbulent viscosity, where $\Omega = \sqrt{G\mstar/r^3}$ is the Keplerian orbital frequency, $c_s$ is the sound speed and \aSS\ is the dimensionless parameter introduced by \cite{ShakuraSunyaev1973} to quantify the turbulent transport of angular momentum. The second term describes the angular momentum extracted vertically by the magnetic disk wind and the third term gives the mass loss induced by the disk wind.

Along the same lines as \aSS, the dimensionless parameter \aDW\ has been introduced to quantify the wind torque exerted on the disk, defined such that the local accretion rate driven by the wind can be written as
\begin{equation}
\label{eq: wind accretion rate}
\dot{M}_{\rm acc}^{DW}(r) = \frac{3\pi\Sigma c_s^2 \aDW}{\Omega}.
\end{equation}
To first order, \aDW\ is proportional to the disk magnetization, that is the ratio between magnetic and thermal pressure.
In this framework the ratio of the local mass accretion rates driven by turbulence and MHD disk winds is approximately equivalent to the ratio between their respective $\alpha-$parameters $\dot{M}_{\rm acc}^{DW}(r)/\dot{M}_{\rm acc}^{visc}(r)\approx \aDW/\aSS$ (see \citealt{Tabone2021a}).

Finally the magnetic lever arm parameter $\lambda$ is introduced
\begin{equation}
\label{eq: lever arm}
\lambda \equiv \frac{L}{\ro\Omega(\ro)},
\end{equation}
where $L$ is the total specific angular momentum extracted by the disk wind streamline launched from \ro\ (see, e.g. \citealt{BlandfordPayne1982}). Using this definition and the conservation of angular momentum the wind-driven mass loss rate $\dot{\Sigma}_W$ is given by (see Eq. 9 in \citealt{Tabone2021a})
\begin{equation}
\label{eq: mass loss rate}
\dot{\Sigma}_W(r) = \frac{3\aDW c_s^2}{4\left(\lambda - 1\right)\Omega r^2} = \frac{1}{2\left(\lambda -1\right)}\frac{\dot{M}_{\rm acc}^{DW}(r)}{4\pi r^2}
\end{equation}

\subsection{Analytical solutions for the disk surface density}
\label{sec: surface density profile}

Equation \ref{eq: master equation} shows the time evolution of the surface density in its most general form where both viscous evolution and wind-driven evolution are considered. 
In this work we will focus on the pure disk wind evolution by setting \aSS\ to zero. For simplicity we will also assume that $\lambda$ is constant. 

For \aDW\ we will consider two prescriptions. 
The first is that \aDW\ is constant with radius and time, similar to what is commonly assumed for \aSS. This simple model, which we will refer to as our fiducial model, captures the main features of wind-driven disk evolution.
The second is that \aDW\ is constant with radius, but scales with the characteristic surface density as $\aDW(t)\propto\Sigma_c^{-\omega} (t)$. This prescription makes \aDW\ time-dependent, as the surface density decreases through mass accretion onto the star and mass loss due to the disk wind.
The prescription mimics the time evolution of the magnetic field strength (see Section 3.4 in \citealt{Tabone2021a}).
It can be shown that for this definition of \aDW\ the disk fully dissipates in a finite time $t_{\rm disp} = 2\tacc/\omega$ (see \citealt{Armitage2013,Tabone2021a}). \cite{Tabone2021b} showed that this prescription of \aDW\ can reproduce observed stellar accretion rates and disk dispersal timescales.

In either case a self-similar solution of Eq. \eqref{eq: master equation} for a disk of finite size takes the following form (see Appendix C in \citealt{Tabone2021a})
\footnote{Note that self-similar solutions can also be found for $\alpha-$parameters that vary as a power-law with radius. For a full discussion, see \citealt{Tabone2021a}.}

\begin{align}
\label{eq: surface density general}
\Sigma(r,t) &= \Gamma(\xi + 1) \Sigma_{\rm c}(t) \left( \frac{r}{\ro}\right)^{-1+\xi} \exp \left[-\frac{r}{\ro}\right] \\
            &= \Gamma(\xi + 1) \frac{\mdisk(t)}{2\pi\ro^2} \left( \frac{r}{\ro}\right)^{-1+\xi} \exp \left[-\frac{r}{\ro}\right] \\
\end{align}
Here $\Sigma_{\rm c}(t)$ and $\mdisk(t)$ are the time-dependent characteristic surface density and disk mass respectively, \ro\ is the characteristic disk size, which does not change with time for wind-driven disk evolution, and $\Gamma$ is the gamma function. The parameter $\xi$ is the mass ejection index \citep{FerreiraPelletier1995,Ferreira1997}
\begin{equation}
\label{eq: xi parameter}
\xi \equiv \frac{\mathrm{d}\ln\dot{M}_{\rm acc}}{\mathrm{d}\ln r} = \frac{1}{2\left(\lambda-1\right)}.
\end{equation}
Note that $\xi>0$, meaning that due to the mass loss caused by the MHD disk wind the disk will have a flatter slope of the surface density $(\Sigma\propto r^{-1+\xi})$ compared to a purely viscously evolving disk.

\begin{figure}
    \centering
    \includegraphics[width=\columnwidth]{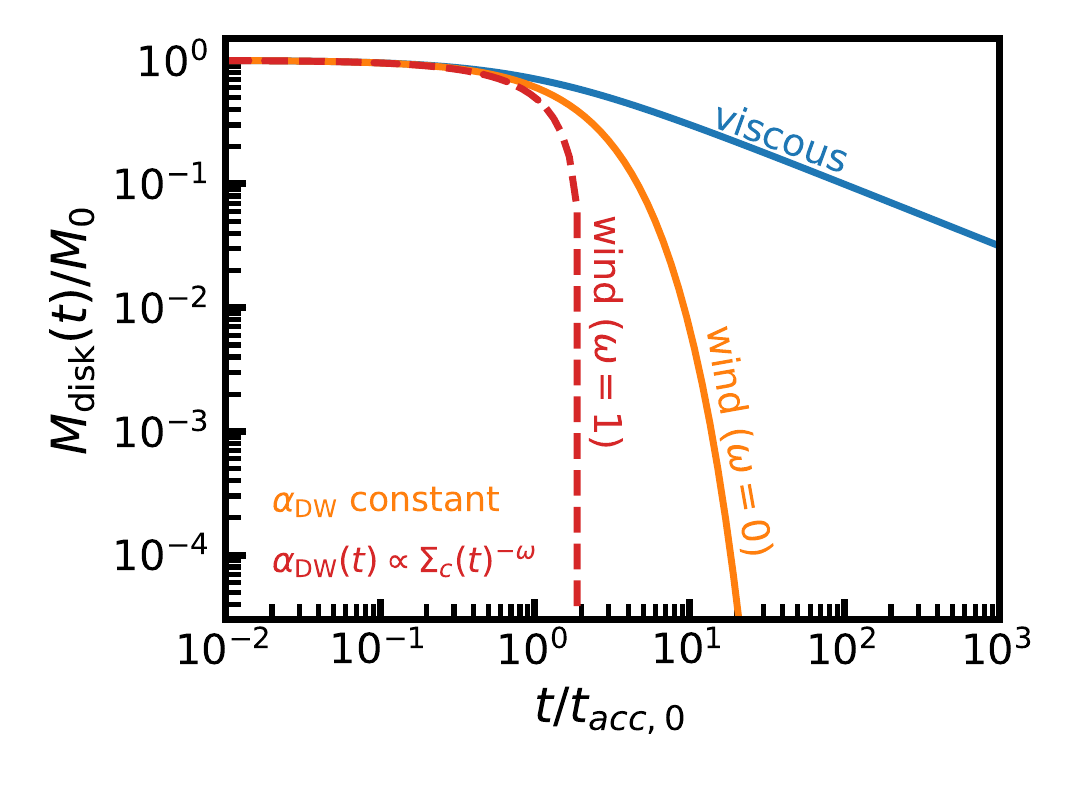}
    \caption{\label{fig: disk mass evolution} Evolution of the disk mass for different disk evolution scenarios. The solid blue line shows viscous evolution and the orange line shows disk wind evolution with constant \aDW. The dashed line shows the scenario where \aDW\ is time-dependent ($\aDW(t) \propto \Sigma_c(t)^{-\omega}$; see Section \ref{sec: surface density profile}). Note that in this case the disk fully dissipates in a finite time.}
\end{figure}

For our two prescriptions of \aDW\ the time evolution of the disk mass takes the following form
\begin{align}
\label{eq: mass evolution}
\mdisk(t) &= \mo\exp^{\frac{t}{2\tacc}}  &[\aDW\ \mathrm{cst}]\\
\mdisk(t) &= \mo\left(1 - \omega\frac{t}{2\tacc}\right)^{\frac{1}{\omega}} &[\aDW(t)\propto\Sigma_{\rm c}(t)^{-\omega}].
\end{align}
Here \mo\ is the initial disk mass. Note that for a purely wind-driven disk, i.e., where $\aSS = 0$, \ro\ does not change with time, meaning the disk mass is the only disk parameter that evolves.
A comparison of the disk mass evolution for the different models discussed is shown in Figure \ref{fig: disk mass evolution}.
Full derivations can be found in \cite{Tabone2021a}. The evolution is controlled by the accretion timescale
\begin{equation}
\label{eq: accretion timescale}
\tacc\equiv \frac{\ro}{3\epsilon_0 c_s(\ro)^2 \aDW},
\end{equation}
where $\epsilon_0\equiv H_0/\ro$ is the disk aspect ratio at \ro.

\subsection{Initial conditions of the models}
\label{sec: initial condition}

The evolution of the surface density of a wind-driven disk is described by four parameters: the initial disk mass \mo, the initial disk size \ro, the magnetic lever arm $\lambda$ which sets the slope of the surface density through the $\xi$ parameter, and the accretion timescale \tacc\ (Equation \eqref{eq: accretion timescale} implies we could use instead any two parameters from $\{\ro, \aDW, \tacc\}$; here we chose to use \tacc\ and \ro). 

As will be discussed in Section \ref{sec: caveats}, $\lambda$ has only a small effect on the measured outer radius. We therefore fix $\lambda=3$ for this work, which corresponds to $\xi=0.25$.
Similarly we set the accretion timescale of all models to $t_{\rm acc} = 5\times10^5\ \mathrm{yr}$.
As can be seen in Eqs. \eqref{eq: surface density general} and \eqref{eq: mass evolution} the evolution of the surface density is given in terms of $t/\tacc$, meaning that our choice of \tacc\ does not matter if our results are discussed in terms of the dimensionless time $t/\tacc$. However, this becomes difficult when we include the thermochemical model in our analysis, as in that case the time $t$ also enters in the computation of the chemistry (see Section \ref{sec: DALI models}). See Section \ref{sec: caveats} for a discussion to what extent our results are affected by our choice of \tacc.

This leaves two free parameters: \mo\ and \ro. 
For these parameters we explore a range of values to study how the initial conditions affect the evolution of measured gas disk sizes for a wind-driven disk. In particular, we use $\mo\in[10^{-4}, 10^{-3}, 10^{-2}, 10^{-1}]\ \msun$ and $\ro\in[5, 20, 40, 65, 90]\ \mathrm{AU}$, giving us a total of 20 different combinations of initial conditions for our models.

\begin{table}[tbh]
  \centering   
  \caption{\label{tab: model fixed parameters}Fixed \texttt{DALI} parameters of the physical model.}
  \begin{tabular*}{0.8\columnwidth}{ll}
    \hline\hline
    Parameter & Range\\
    \hline
     \textit{Chemistry}&\\
     Chemical age & 0.1-10$^{*,\dagger}$ Myr\\
     {[C]/[H]} & $1.35\cdot10^{-4}$\\
     {[O]/[H]} & $2.88\cdot10^{-4}$\\
     \textit{Physical structure} &\\ 
     $\gamma^{\ddag}$ &  1.0\\
     $\xi^{\ddag}$ & -0.25\\
     $\psi$ & 0.15\\ 
     $h_c$ &  0.1 \\ 
     \ro & [\,5,\,20,\,40,\,65,\,90\,] AU\\
     $M_{\mathrm{gas}}$ & $10^{-7} - 10^{-1,\dagger}$ M$_{\odot}$ \\
     Gas-to-dust ratio & 100 \\
     \textit{Dust properties} & \\
     $f_{\mathrm{large}}$ & 0.9 \\
     $\chi$ & 0.2 \\
     composition & standard ISM$^{1}$\\
     \textit{Stellar spectrum} & \\
     $T_{\rm eff}$ & 4000 K + Accretion UV \\
     $L_{*}$ & 0.28 L$_{\odot}$  \\
     $\zeta_{\rm cr}$ & $10^{-17}\ \mathrm{s}^{-1}$\\
     \textit{Observational geometry}&\\
     $i$ & 0$^{\circ}$ \\
     PA & 0$^{\circ}$ \\
     $d$ & 150 pc\\
    \hline
  \end{tabular*}
  \begin{minipage}{0.75\columnwidth}
  \vspace{0.1cm}
  {\footnotesize{$^*$The age of the disk is taken into account when running the time-dependent chemistry. $^{\dagger}$These parameters evolve with time.
  $^{\ddag}$ Note that for a wind-driven disk the slope of the surface density is given by $\Sigma\propto r^{-\gamma + \xi}$ (see Eq. \ref{eq: surface density general}).
  $^{1}$\citealt{WeingartnerDraine2001}, see also Section 2.5 in \citealt{Facchini2017}. }}
  \end{minipage}
\end{table}

\subsection{The thermochemical code {\normalfont \texttt{DALI}}}
\label{sec: DALI models}

In order to link the evolution of the surface density described by equation \eqref{eq: surface density general} to  resulting evolution of the observed gas disk size we use the thermochemical code \underline{D}ust \underline{A}nd \underline{Li}nes (DALI; \citealt{Bruderer2012,Bruderer2013}). The setup used here is very similar to the approach used in \cite{Trapman2020} to study the effect of viscous evolution on the observed gas disk size. 

For each combination of initial disk mass, \mo, and size, \ro, we calculate the current surface density at 10 consecutive disk ages between 0.1 and 10 Myr using Eq. \eqref{eq: surface density general}. These are then used as input for DALI. For each model, DALI first solves the continuum radiative transfer equation using a Monte-Carlo method to determine both the dust temperature and the radiation field at each point in the disk. Next, the code computes atomic and molecular abundances by solving the time-dependent chemistry for the same disk age that is used to calculate the surface density. The atomic and molecular excitation levels are then determined using a non-LTE calculation. The code then calculates the gas temperature by balancing heating and cooling processes. Because of their interdependence, the chemistry, excitation, and heating and cooling are computed iteratively until a self-consistent solution is found. As a final step the model is ray-traced to produce synthetic emission maps of the $^{12}$CO $J=2-1$ line. A more detailed description of the code can be found in Appendix A of \cite{Bruderer2012}. 

\begin{figure*}[htb]
    \centering
    \includegraphics[width=\textwidth]{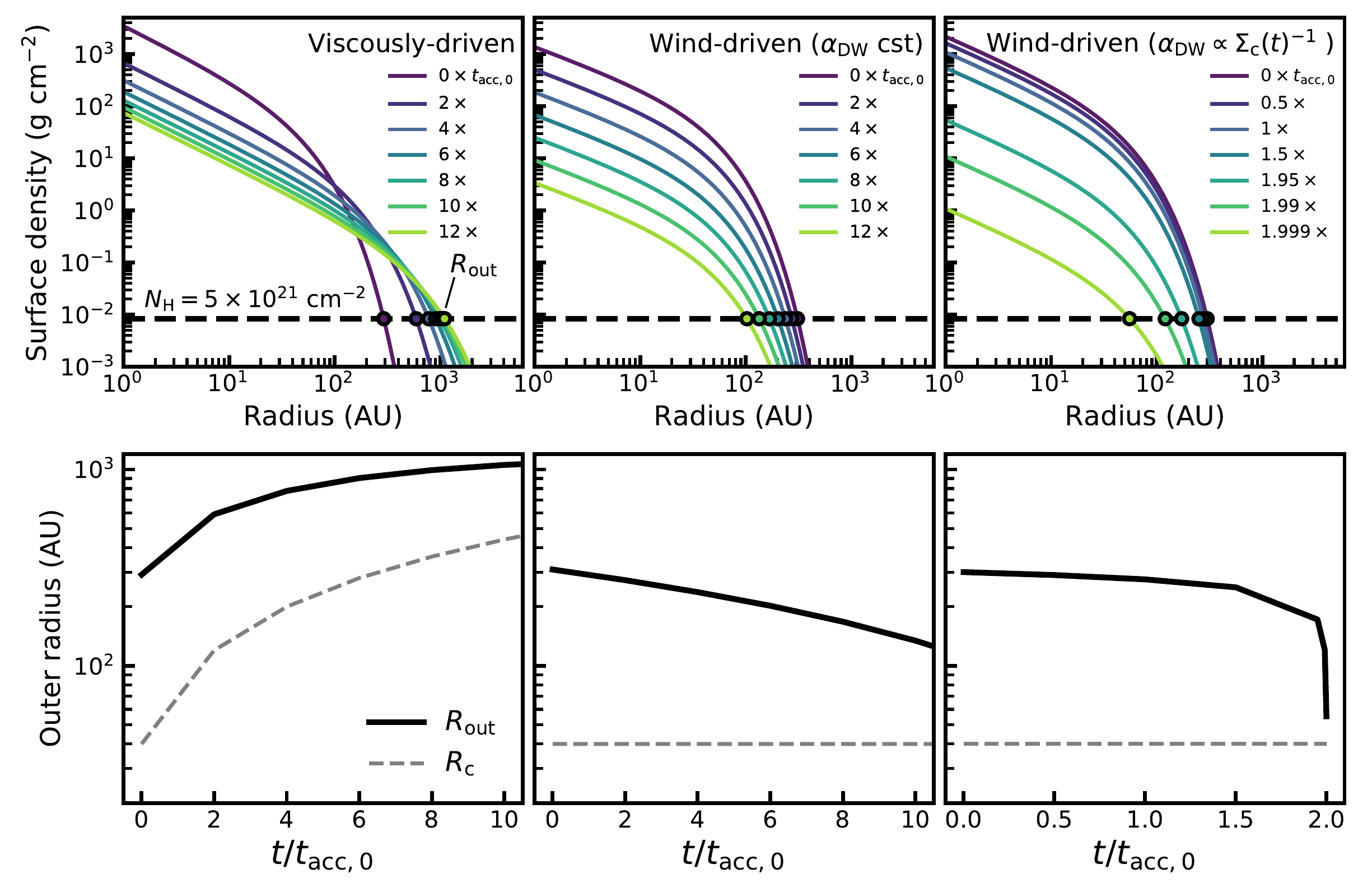}
    \caption{\label{fig: toy example}\textbf{Top panels}: Time evolution of the surface density for three different evolutionary cases: a viscously evolving disk (left), a wind-driven disk with constant \aDW\ (middle) and a wind-driven disk with a time-dependent $\aDW(t)\propto\Sigma_{\rm c}(t)^{-1}$. Colors show different times in the disk evolution. Note that for the time-dependent \aDW\ different timesteps have been chosen to highlight the effect of disk dispersal. The black dashed line denotes the column density cut used to calculate $R_{\rm out}$, which is shown in the bottom panels (see also Section \ref{sec: toy example}). \textbf{Bottom panels:} Time evolution of $R_{\rm c}$ (gray) and $R_{\rm out}$ (black) for the three evolutionary cases shown in the top panels. }
\end{figure*}

To focus on how the wind-driven evolution of the surface density affect the observed gas disk size, we fix the other parameters of the models that, for example, describe the vertical structure of the disk or the distribution of large dust grains. To facilitate the comparison with viscous evolution, we adopt the same parameters as used in \cite{Trapman2020} (see Table \ref{tab: model fixed parameters}).

For the vertical structure, the disk is assumed be to vertically isothermal and in hydrostatic equilibrium, resulting in a Gaussian vertical density structure (e.g. \citealt{ChiangGoldreich1997})

\begin{equation}
\rho_{\rm gas}(R,z) = \frac{\Sigma_{\rm gas}(R)}{\sqrt{2\pi}Rh}\exp\left[-\frac{1}{2}\left(\frac{z/R}{h}\right)^2 \right], 
\end{equation}
where $h$ is the scale height of the disk, parameterized by a powerlaw $h = h_c (R/R_{\rm c})^{\psi}$ to include disk flaring.

The settling of large dust grains is included by splitting the dust grains into two populations, following \citealt{Andrews2011}. Small grains (0.005\,-\,1\,$\mu$m) are included with a mass fraction $1-f_{\rm large}$ and are distributed vertically in the same manner as the gas. Large grains (1\,-\,$10^3$\,$\mu$m) make up the remainder of the dust mass $(f_{\rm large}\times\mdust)$. These grains are limited to a vertical region with a scale height $\chi h$ (\,$\chi<1$\,) to simulate the effect of dust settling.

Finally the stellar spectrum is a black body with an effective temperature of $T_{\rm eff} = 4000\,$K. The spectrum was scaled to a stellar luminosity of $L_* = 0.28\,\mathrm{L}_{\odot}$. Excess ultraviolet (UV) radiation, in the form of a 10000\,K blackbody, was added to the spectrum to account for stellar accretion. The luminosity of this component was determined by taking a stellar mass accretion rate of $10^{-8}\ \mathrm{M}_{\odot}\ \mathrm{yr}^{-1}$ and assuming that the gravitational potential energy of the accreted mass is released with 100\,\% efficiency (see, e.g. \citealt{kama2015}). For our models this results in a total far ultraviolet (FUV) luminosity of $2.7\times10^{31}\ \mathrm{erg\ s^{-1}}$ between 0.0911 and 0.206 micron.

We note that we do not include the evolution of stellar mass accretion rate as a result of our disk evolution. Our tests show that the effect of the stellar FUV luminosity on observed outer radius is minimal. This is because in the outer disk, the region most important for the observed outer radius, the UV radiation is dominated by the interstellar radiation field.
For completeness, we have also tested the effect of the external UV field on the observed outer radius. We find that an increase from 1 G$_0$ to 30 G$_0$ decreases the observed outer radius by less than 40\%. We note however that this not include the effect of external radiation on the density structure in the outer disk.
All assumed parameters are summarized in Table \ref{tab: model fixed parameters}.

\section{Results}
\label{sec: results}

\subsection{Viscous versus wind-driven evolution: a toy model}
\label{sec: toy example}

Before moving to the DALI models, let us first compare the size evolution of a wind-driven disk to that of a viscously evolving disk using a simplified toy model. The size of the gas disk is commonly measured from $^{12}$CO rotational emission (see, e.g. \citealt{barenfeld2017,ansdell2018,Sanchis2021}). For observations with a high enough sensitivity the observed outer edge of the disk would be the point in the outer disk where the CO column density drops below the threshold where it is able to effectively self-shield against photo-dissociation by FUV photons (see, e.g. \citealt{vanDishoeckBlack1988}). Assuming that the vertically averaged CO abundance is approximately constant, this would imply that the observed outer radius lies at a fixed surface density. We will revisit this approach in more detail in a forthcoming paper.

For now, to get a feeling for how the observed size of the disk evolves over time for different evolution scenarios, we define the observed outer radius as the radius where the surface density drops below $5\times10^{21}\ \mathrm{cm}^{-2}$ and examine how this radius evolves over time for a viscously evolving disk and two wind-driven evolving disks: one with a constant \aDW\ and one with a time-dependent $\aDW (t) \propto \Sigma_{\rm c}(t)^{-1}$.

The top panels of Figure \ref{fig: toy example} shows how the surface density evolves for these three scenarios. Each model starts with $\mo \equiv \mdisk(t=0)=0.1\ \msun$ and $\ro \equiv \rc(t=0) = 40\ \mathrm{AU}$. The black points mark the location of the observed gas outer radius \rout\ using our surface density-based definition. The bottom panels show, for each evolutionary scenario, how \rout\ and the characteristic radius \rc\ change over time. For the viscously evolving disk both \rout\ and \rc\ increase with time, a process commonly referred to as viscous spreading.


\begin{figure*}[thb]
    \centering
    \begin{minipage}{0.95\textwidth}
    \includegraphics[width=\textwidth]{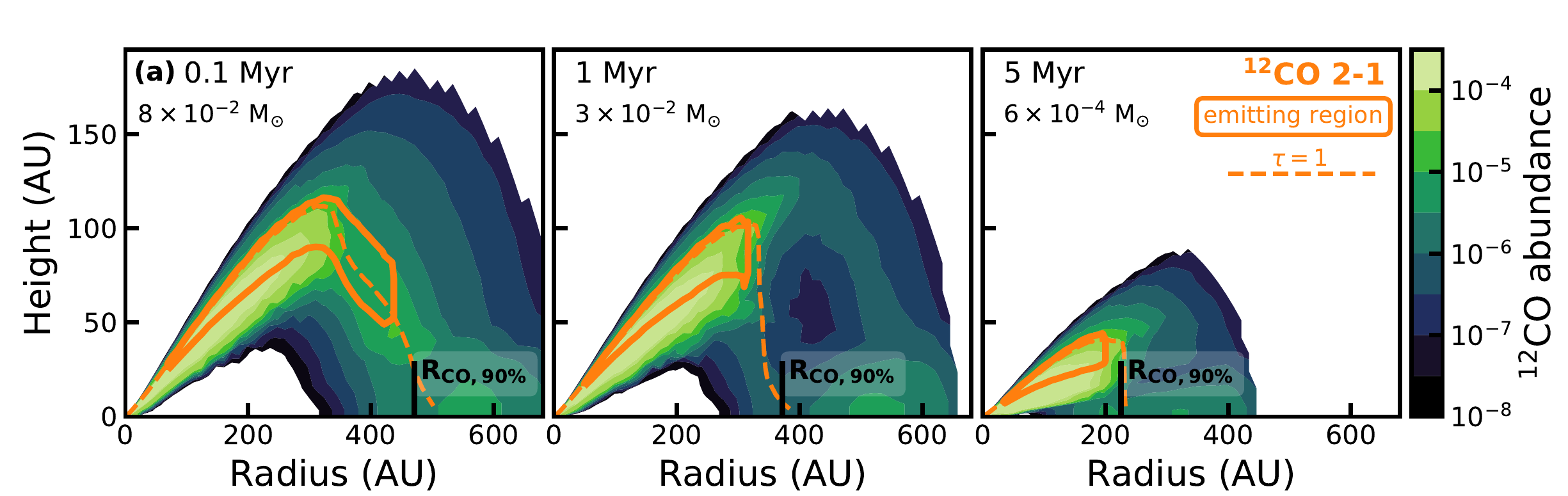}
    \end{minipage}
    \begin{minipage}{0.51\textwidth}
    \includegraphics[width=\textwidth]{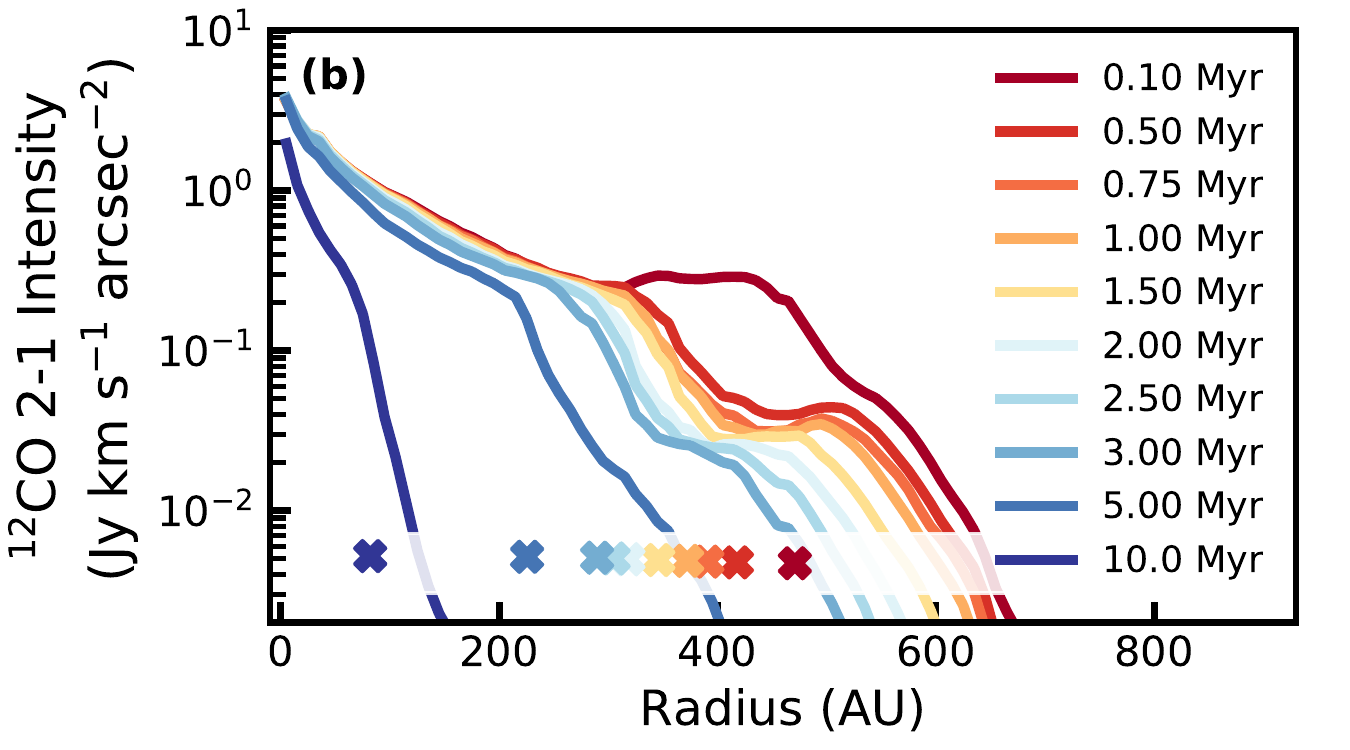}
    \end{minipage}
    \hspace{0.25cm}
    \begin{minipage}{0.455\textwidth}
    \includegraphics[width=\textwidth]{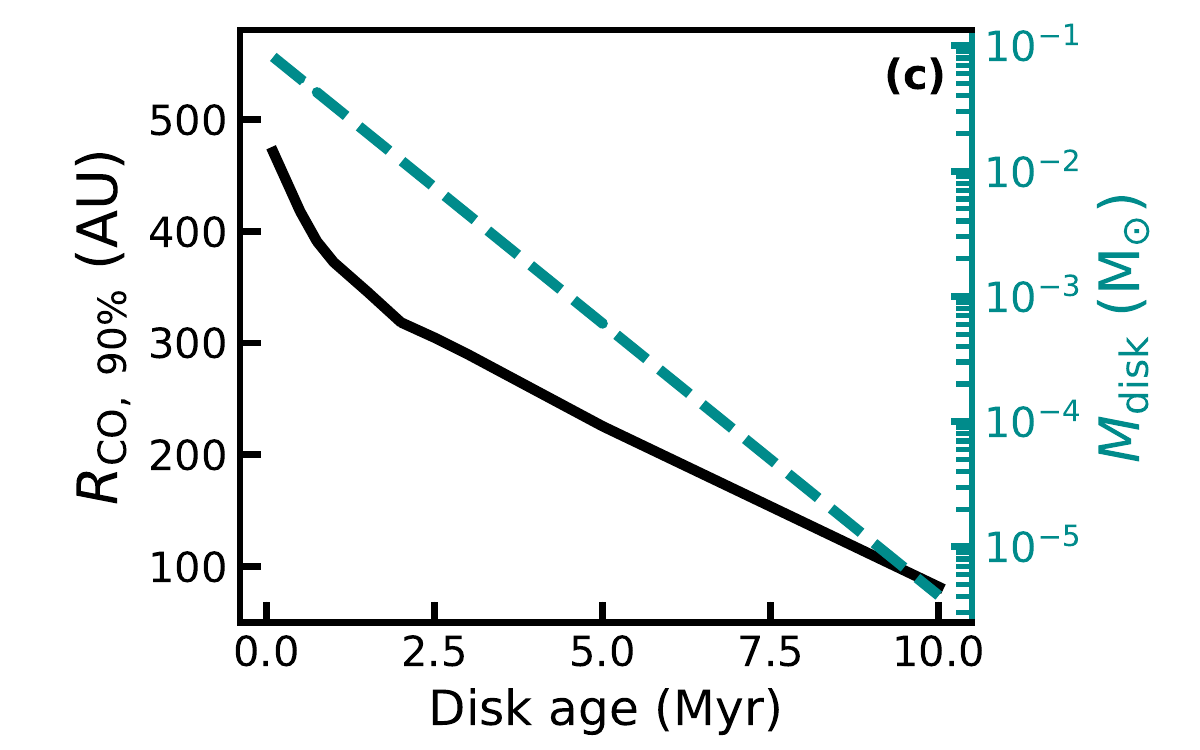}
    \end{minipage}
    \caption{\label{fig: time evolution example}
    \textbf{Top panels}: CO abundance distribution for a wind-driven disk model with $\mo = 0.1\ \msun,\ \ro = 65$ AU, shown for three different time steps. Shown in orange are the emitting region that encloses 75\% of the total flux (solid contour) and $\tau=1$ surface (dashed line) for the $^{12}$CO 2-1 emission line. The radius enclosing 90\% of the CO 2-1 emission is marked by a black vertical line. 
    \textbf{Bottom left panel}: time evolution of CO 2-1 intensity profile for the model shown in the top panels. Here the various colors indicate different disk ages between 0.1 and 10 Myr $(0.2-20\times\tacc)$. The crosses at the bottom of the panel mark \rgas, the radius that encloses 90\% of the total CO 2-1 flux. 
    \textbf{Bottom right panel}: \rgas\ versus disk age for the same model shown in the other panels $(\mo=0.1\ \msun,\ro=65\ \mathrm{AU})$. Right hand side shows the evolution of the disk mass of the model.}
\end{figure*}

In contrast, the middle and rightmost panel of Figure \ref{fig: toy example} shows that \rout\ decreases with time for a wind-driven disk. As \ro\ stays constant in this case, the time-evolution of \rout\ for a wind-driven model is solely determined by the evolution of the disk mass. As a result of the decreasing disk mass the radius at which the surface density equals $5\times10^{21}\ \mathrm{cm}^{-2}$ moves inward, meaning \rout\ decreases with time.
For a constant \aDW\ \rout\ decreases steadily over time, driven by the exponential decrease of the disk mass. In contrast, if \aDW\ scales with the characteristic surface density \rout\ does not change significantly over most of the disk lifetime $(t/\tacc \approx 0-1.6)$ until \rout\ rapidly drops as the disk starts to dissipate $(t/\tacc=1.9-2.0)$. This highlights how the evolution of the disk mass can significantly alter the evolution of \rout.

\subsection{Time evolution of the $^{12}$CO emission profile and observed gas outer radius for constant \aDW}
\label{sec: evolution - cst alpha - single model}

We now examine the time evolution of a wind-driven disk using a representative DALI model with a constant \aDW. The model discussed here has an initial disk mass of $\mo=0.1\ \msun$, an initial size of $\ro=65$ AU and an accretion timescale of $\tacc = 0.5$ Myr. 

The top panels of Figure \ref{fig: time evolution example} show the CO abundance structure at three times during the evolution of the disk: 0.1, 1 and 5 Myr, which corresponds to 0.2, 2 and 10 accretion timescales. As time progresses the CO-rich warm molecular layer can be seen moving closer towards the disk midplane. This is not related to a decrease in the height of the disk, which is kept fixed, but rather it is related to the decreasing disk mass. For a disk with a lower disk mass stellar radiation is able to penetrate deeper into the disk, meaning far-ultraviolet photons are able to photodissociate CO at a lower scale height, thus moving the CO-rich layer downward. The stellar radiation also heats up the disk, increasing the temperature around the midplane of the disk and thus reducing how much CO freezes out.
Figure \ref{fig: time evolution example} shows that most of the $^{12}$CO $J=2\,-\,1$ emission originates from the CO-rich molecular layer and is optically thick throughout most of the disk. At approximately $200-300$ AU, depending on the disk mass, the molecular layer ends and the CO column density drops, resulting in the CO 2-1 emission becoming optically thin beyond this radius. The observed outer radius \rgas, defined as the radius that encloses 90~\% of the CO 2-1 flux, lies approximately at this transition point between optically thick and optically thin CO emission. Note that \rgas\ marks the radius in the disk at which CO is largely removed from the gas, which was the assumption we made for our toy model. 

The full evolution of the CO emission is presented in the bottom left panel of Figure \ref{fig: time evolution example}, which shows the CO 2-1 intensity profile at 10 time-steps between 0.1 and 10 Myr (similar profiles for models with different \mo\ and \ro\ are shown in Figure \ref{fig: effect of initial conditions}). The inner $\sim200$ AU correspond to the optically thick portion of the emission.
The transition from optically thick to optically thin emission is also clearly visible as a drop of approximately an order of magnitude in intensity at $\sim300-400$ AU. The optically thin emission beyond $\sim 400$ AU is linked to the $x_{\rm CO}\approx10^{-5}$ gas around the disk midplane that be clearly seen at a radius of $\sim500$ AU in the middle top panel of Figure \ref{fig: time evolution example}. For most of the midplane of the disk CO is frozen out. The increase in CO abundance at the midplane in the outer disk is the result of external radiation photo-desorbing CO from the grains. Compared to the optically thick CO in the inner disk the optically thin CO only makes up a small fraction of the total CO flux and therefore only has a small effect on the location of \rgas. 

The observed outer radii \rgas\ for each of the time-steps is marked by a cross shown in the bottom left panel of Figure \ref{fig: time evolution example}. As discussed previously, \rgas\ approximately coincides with the drop in the intensity profile resulting from the CO emission becoming optically thin. The bottom right panel of Figure \ref{fig: time evolution example} shows that \rgas\ decreases over time. This decrease becomes linear with time for disk ages much larger than the accretion timescale. This gives some credence to our approximation of \rout\ in Section \ref{sec: toy example}. If we assume that \rgas\ is linked to a fixed surface density $\Sigma_{\rm cut}$ and \rgas\ lies in exponential taper, we obtain $\mdisk \exp(-\rgas/\ro) = \mathrm{cst}$ (cf. Eqs. \eqref{eq: surface density general} and \eqref{eq: mass evolution}). This equation can be rewritten to show that 
\begin{equation}
\label{eq: size proportionality}
\rgas \propto \ro\left( \ln\left[\mo\ \ro^{-2}\ \Sigma_{\rm cut}^{-1}\right] - t/2\tacc\right).    
\end{equation}
Note that this example applies for a constant \aDW. We will investigate the relation between \rgas\ and \ro\ in more detail in a forthcoming paper.

\subsection{Effect of the initial disk mass and size}
\label{sec: evolution - mass and size}

Having shown the results for a single model in the previous section, we now investigate the effect of the initial disk mass and size on the evolution of the gas disk size.

\subsubsection{Constant \aDW}
\label{sec: evolution - cst alpha - mass and size}

\begin{figure}
    \centering
    \begin{minipage}{\columnwidth}
    \includegraphics[width=\columnwidth]{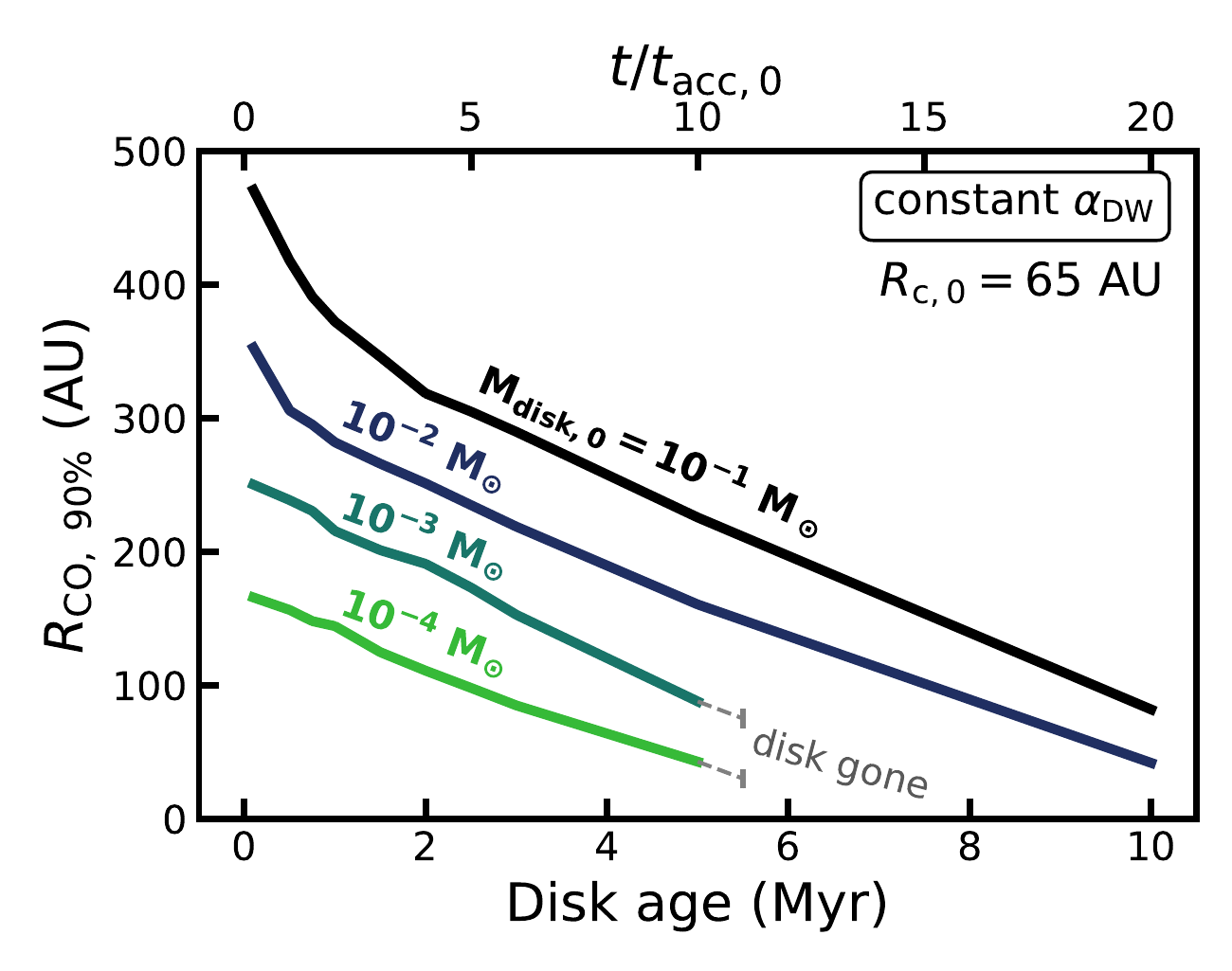}
    \end{minipage}
    \begin{minipage}{\columnwidth}
    \includegraphics[width=\columnwidth]{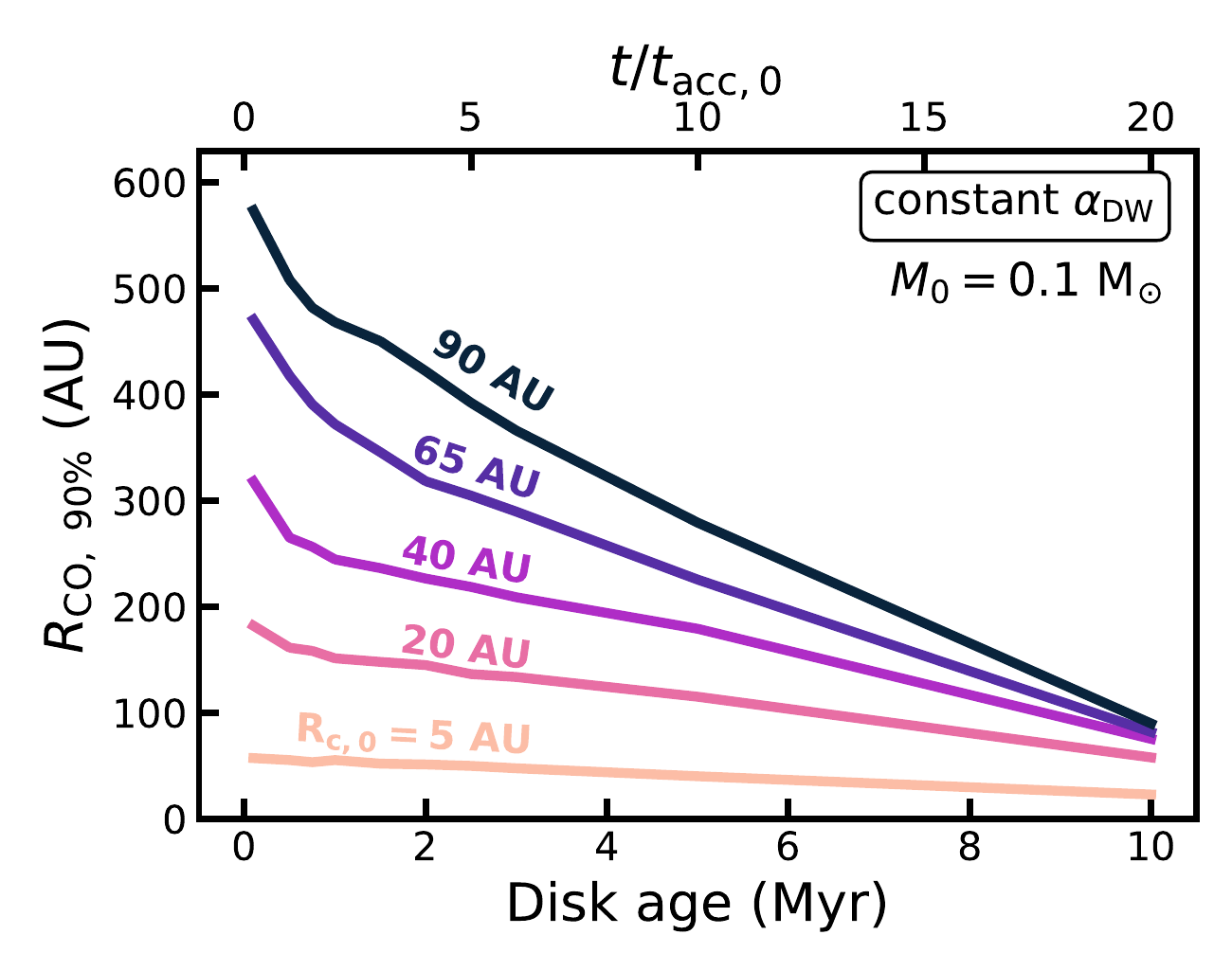}
    \end{minipage}
    \caption{\label{fig: disk size vs time} 
    \rgas\ versus time for constant \aDW\ models with different initial disk mass (top panel) or different initial disk size (bottom panel). All models in the top panel have $\ro=65$ AU and all models in the bottom panel have $\mo=0.1\ \msun$. In the top panel two models are not shown, namely those with $\mo\leq10^{-3}\ \msun$ at $t=10$ Myr. For these disks the CO 2-1 emission has dropped to negligible levels (cf. Figure \ref{fig: effect of initial conditions}). Note that apparent convergence seen in the bottom panel disappears if \rgas\ is expressed in terms of \ro\ (see Figure \ref{fig: normalized disk size vs time} in Appendix \ref{app: Extra figure Rgas/Rc}).
    For reference, the top axis of both panels shows the dimensionless time $t/\tacc$ that goes into the evolution of \mo.  }
\end{figure}

Let us first consider the case where \aDW\ is constant.
Figure \ref{fig: disk size vs time} shows how the initial mass and size affect the way \rgas\ changes over time. As mentioned at the end of Section \ref{sec: evolution - cst alpha - single model}, \rgas\ decreases linearly with time for $t\gg\tacc$. Note that models with $\mo\leq10^{-3}\ \msun$ at 10 Myr are not shown, as their CO 2-1 emission has dropped to negligible levels (cf. Figure \ref{fig: effect of initial conditions}).
The top panel highlights that the \rgas\ is related to the mass of the disk. A more massive disk has a larger CO column further out in the disk, meaning that the CO emission remains optically thick up to larger radii (see also \citealt{Trapman2019a, Trapman2020}). Interestingly, the evolution of \rgas\ for models with different \mo\ is almost parallel, indicating that changing \mo\ does not significantly affects the evolution of \rgas. We note that this is what we would expect based on Equation \eqref{eq: size proportionality}.
The bottom panel of Figure \ref{fig: disk size vs time} shows that models with a different \ro\ are not parallel, with the slope being more steep for a model with a larger \ro. If we divide \rgas\ by \ro\ the models become parallel again (see Figure \ref{fig: normalized disk size vs time}), indicating that $\rgas/\ro \propto -t/2\tacc$ 
(cf. Equation \eqref{eq: size proportionality}).

\subsubsection{$\Sigma_{\rm c}$-dependent \aDW}
\label{sec: evolution - time-dep alpha - mass and size}

\begin{figure}
    \centering
    \includegraphics[width=\columnwidth]{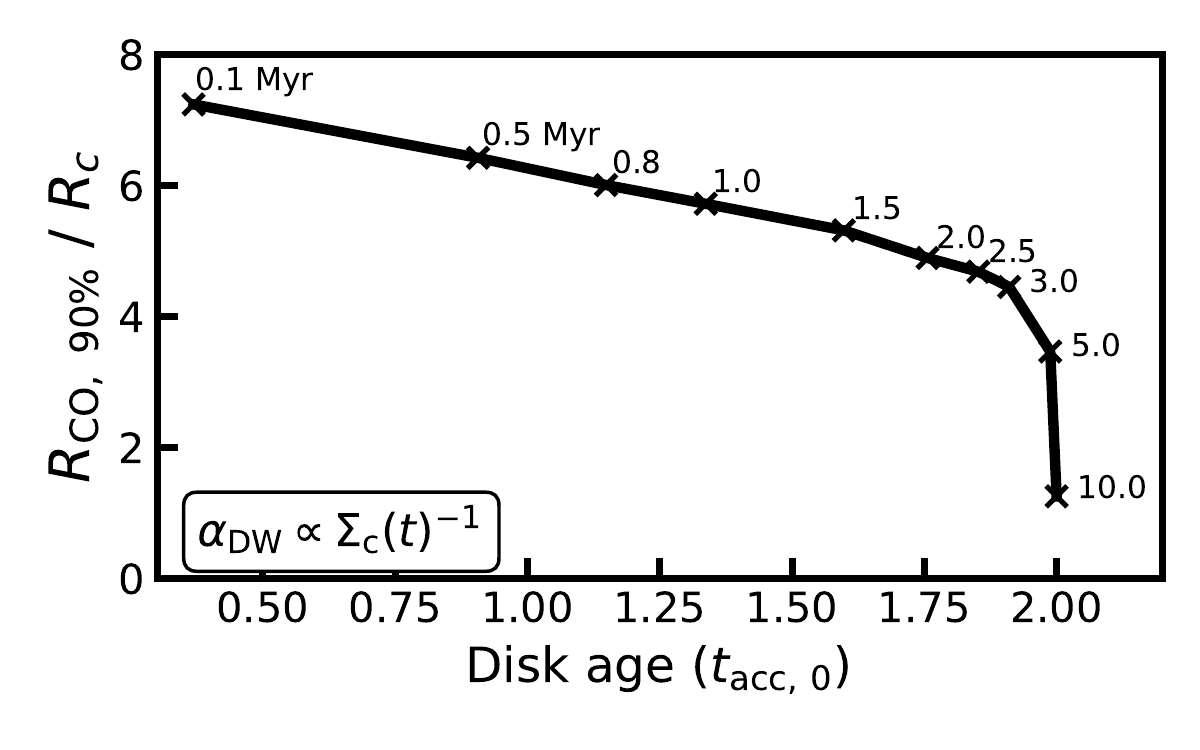}
    \caption{\label{fig: constant B fitting} Example of remapping the mass evolution of constant \aDW\ model to the mass evolution where $\aDW(t)\propto\Sigma_{\rm c}(t)^{-1}$. The crosses show how the \rgas\ from the original timesteps were remapped to the disk mass evolution as a result of a time-dependent \aDW\ (see Section \ref{sec: surface density profile}).}
\end{figure}

In Section \ref{sec: toy example} we saw that changing the evolution of the disk mass can have a drastic effect on the the time evolution of \rgas. Here we move from a constant \aDW, where the disk mass decreases exponentially, to assuming that $\aDW\propto\Sigma_c(t)^{-\omega}$, where $\Sigma_c$ is the surface density at \ro.
For simplicity we only examine $\omega=1$, which corresponds to a constant magnetic flux. In this case the disk mass decreases linearly with time (see Eq. \eqref{eq: mass evolution}).

\begin{figure}
    \centering
    \begin{minipage}{\columnwidth}
    \includegraphics[width=\columnwidth]{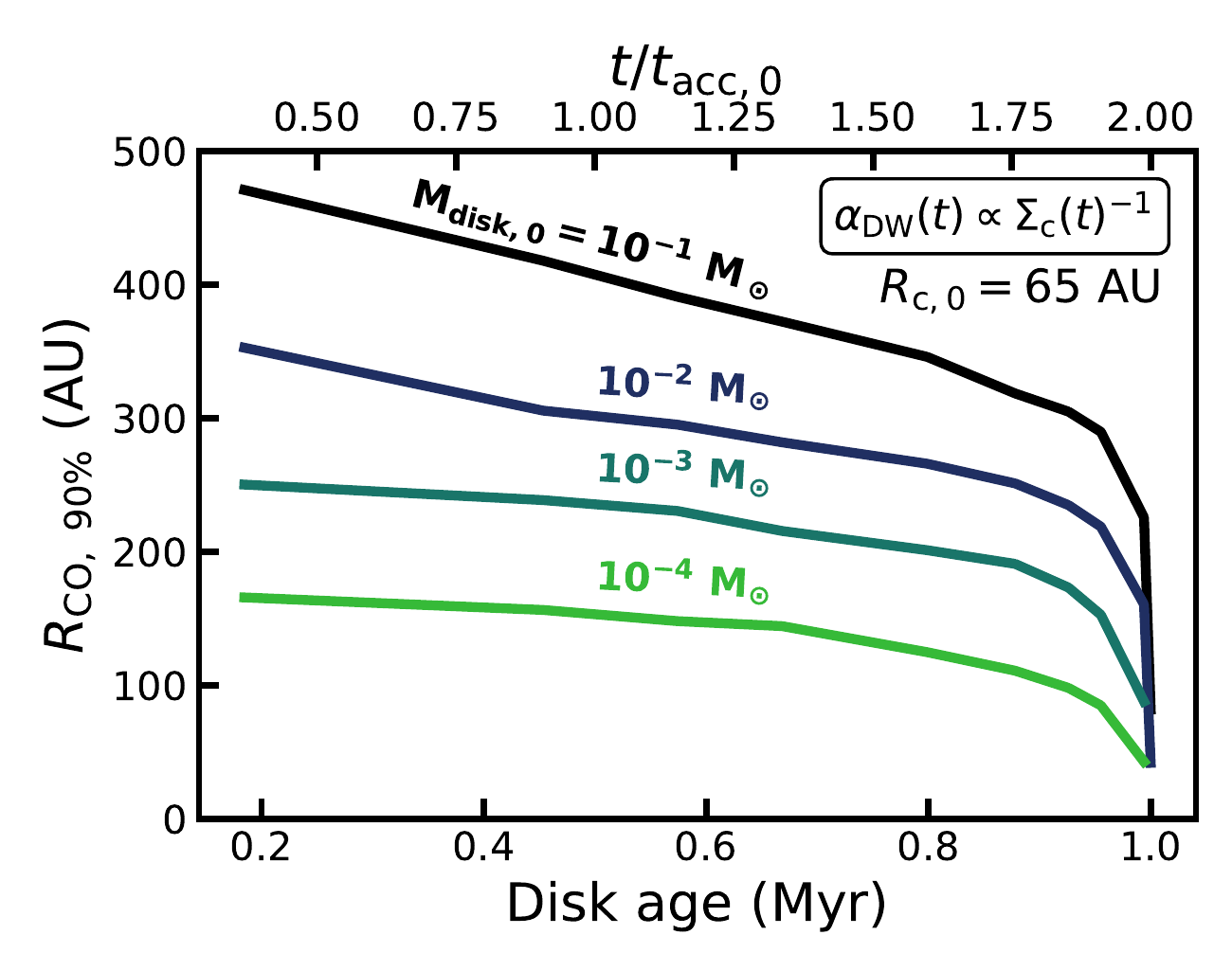}
    \end{minipage}
    \begin{minipage}{\columnwidth}
    \includegraphics[width=\columnwidth]{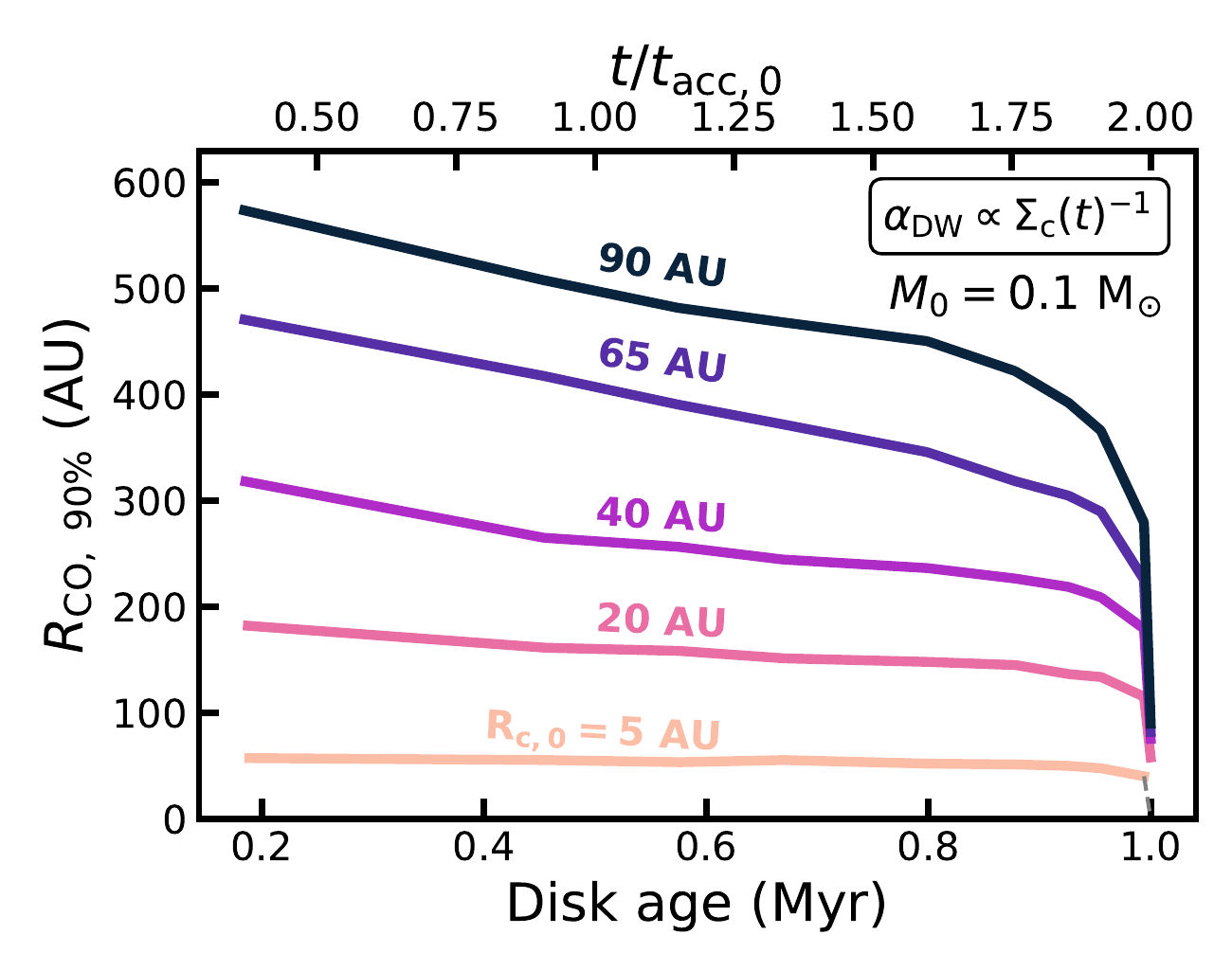}
    \end{minipage}
    \caption{\label{fig: disk size vs time sigma alpha} 
    As Figure \ref{fig: disk size vs time}, but showing the evolution for $\alpha_{\rm DW}(t)\propto\Sigma_{\rm c}(t)^{-1}$. }
\end{figure}

Rather than running a new set of models to investigate this, 
we using our existing models, take the disk mass at each of the timesteps and calculate, using Equation \eqref{eq: mass evolution}, at which time $t/\tacc$ the disk would have this disk mass, given the new disk mass evolution (See Figure \ref{fig: constant B fitting}). Note that this will result in the chemical age, i.e. the time used in the time-dependent chemistry in DALI, being different from the age of the disk. However, we will show in Section \ref{sec: caveats} that this only has a minimal impact on \rgas.

Analogous to Figure \ref{fig: disk size vs time}, Figure \ref{fig: disk size vs time sigma alpha} shows the evolution of \rgas\ for a range of initial disk masses and sizes for the case where $\aDW(t)\propto\Sigma_{\rm c}(t)^{-1}$. Similar to Figure \ref{fig: toy example} \rgas\ decreases slowly for most of the disk lifetime. However, at $\sim1.85\times\tacc$ \rgas\ suddenly decreases rapidly until the disk is fully dispersed at $t = 2\times\tacc$. 
Increasing \mo\ and \ro\ both increase the initial \rgas. Both also slightly increase the rate at which \rgas\ decreases, but the predominant feature remains the sudden drop in \rgas\ as the disk disperses. 
The observational implication of this would be that disks of different ages will have very similar disk sizes until they rapidly disappear as their age reaches two times their accretion timescale. In other words, the size distribution of a population of disk will be set by the initial size and mass distributions and will not significantly evolve over time, apart from disks disappearing from the population as they disperse. 

\section{Discussion}
\label{sec: discussion}

\subsection{Comparing to observations}
\label{sec: comparing to observations}

\begin{figure*}
    \centering
    \includegraphics[width=\textwidth]{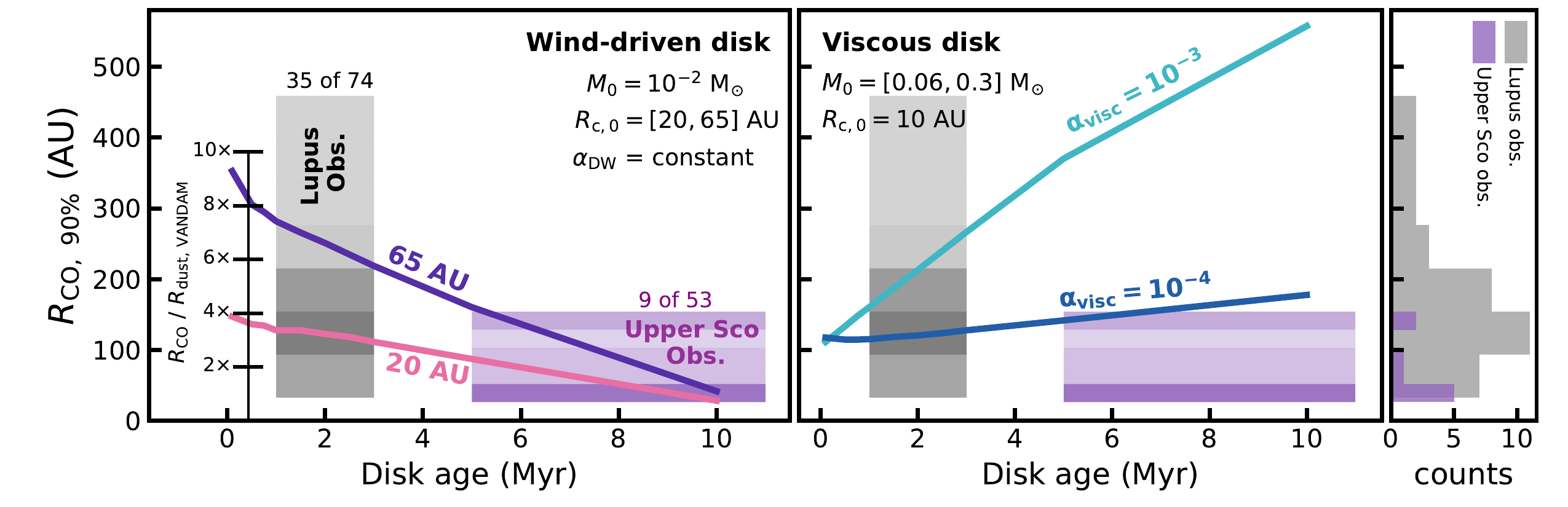}
    \caption{\label{fig: comparing to observations}
    Comparison between gas disk sizes of a wind-driven disk model (left panel), viscously evolving disk models (middle panel) and observations in Lupus (gray) and Upper Sco (purple). See Section \ref{sec: comparing to observations} for a detailed description of the observations. Viscously evolving disk models were obtained from \cite{Trapman2020}. Shown here are the models with $\mstar = 1.0\ \msun$ and $\alpha_{\rm visc} = 10^{-3}, 10^{-4}$.  Histograms of both sets of observations are shown in the rightmost panel. Top-down views of these histograms are included in the left and middle panel, where a darker shade in a given bin corresponds to a higher count in said bin. The x-axis location and width of these histograms corresponds to the age range of both star-forming regions.
    The ticked vertical line shows the mean $R_{\rm dust,\ VANDAM}$ from the VANDAM survey presented in \cite{Tobin2020}, times a multiplier denoted by the horizontal line. For example, the line at '$4\times$' would show the mean \rgas, assuming $\rgas/R_{\rm dust,\ VANDAM} = 4$. The width of the horizontal lines is the average age range of Class I sources (see, e.g. \citealt{Evans2009}).
    }
\end{figure*}

There are currently few large samples of protoplanetary disks for which the gas disk size has been measured in a homogeneous manner. Disks in the Lupus star-forming region have been observed with ALMA by \cite{ansdell2018} and \cite{Sanchis2021}. Of the 74 disks detected in the millimeter continuum 51 were also detected in $^{12}$CO 2-1. \cite{ansdell2018} measured the gas disk size from the $^{12}$CO emission for 22 disks $(30\%)$, but due to the low signal-to-noise of the observations they were not able to measure gas disk sizes for the remaining 29 disks. Recently \cite{Sanchis2021} were able to increase the number of measured gas disk sizes to 35 disks $(43\%)$ in Lupus through a careful analysis and fitting of the $^{12}$CO emission.

\begin{figure}[ht]
    \centering
    \includegraphics[width=\columnwidth]{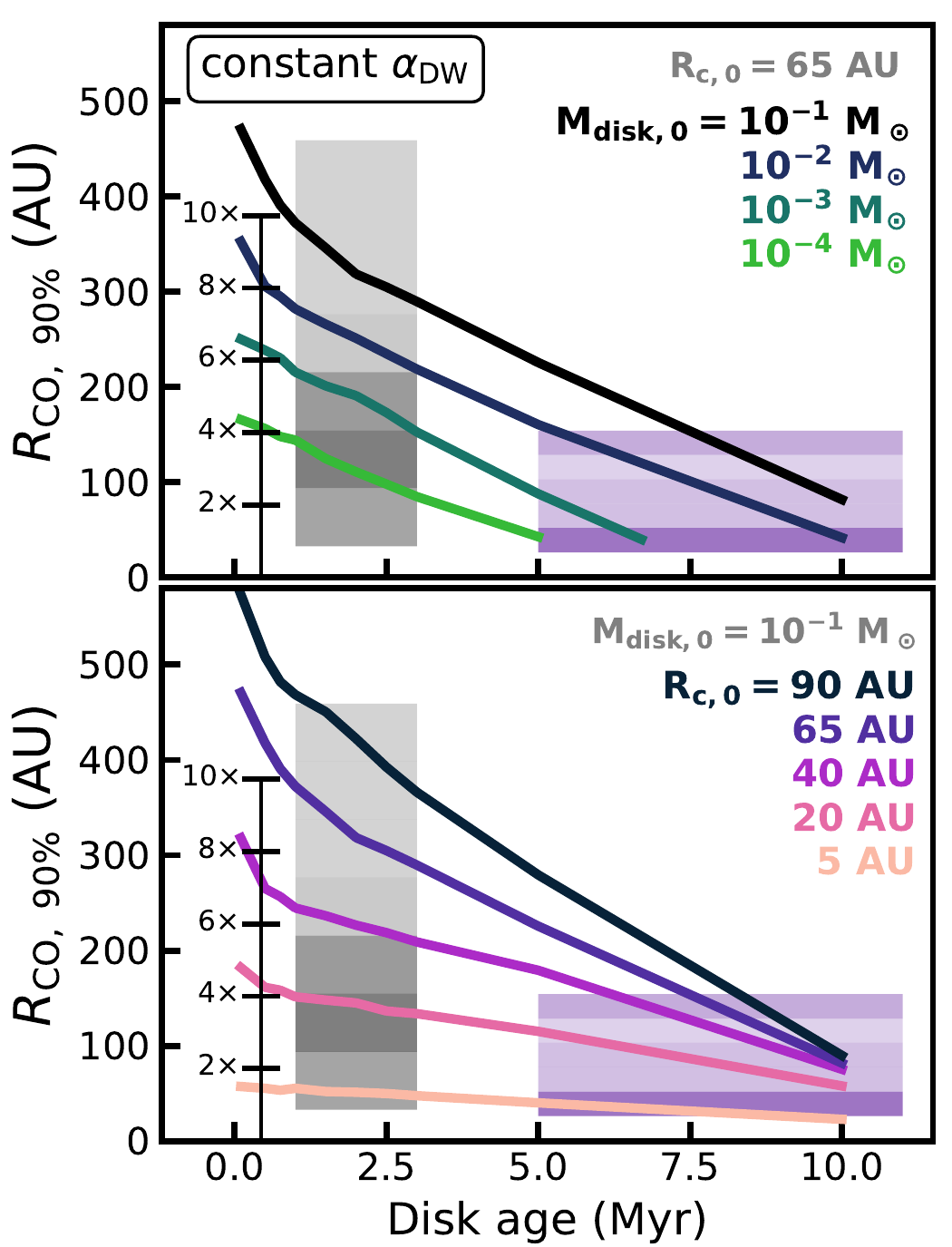}
    \caption{\label{fig: comparing initial conditions to observations}
    Example of the possible range of initial disk sizes and masses of a wind-driven disk that can explain both the observations in Lupus and Upper Sco. The top panel shows that initial disk masses between $10^{-3}\ \msun \leq \mo \leq 0.1\ \msun$ can explain both sets of observations. Note that for the models in the top panel the initial disk size is fixed at $\ro = 65$ AU. Similarly, the bottom panel shows that an initial disk size of $\ro=65$ AU matches the top end of the observed distribution in both Lupus and Upper Sco, whereas the an initial disk size of $\ro = 20$ AU matches the peak of both distributions. It should be noted here that in Upper Sco \rgas\ is measured for only $\sim17\%$ of disks detected in millimeter continuum.
    The ticked vertical line shows the mean Gaussian $2\sigma$ radius $(R_{\rm dust,\ VANDAM})$ from the VANDAM survey presented in \cite{Tobin2020}, times a multiplier denoted by the horizontal line (cf. Figure \ref{fig: comparing to observations}).
    }
\end{figure}

The other region for which gas disks sizes have been measured is Upper Sco. \cite{barenfeld2017} presented ALMA observations of $^{12}$CO 3-2, which was detected for 21 out of the 53 protoplanetary disks detected in the continuum. They modeled the $^{12}$CO visibilities, but due to low signal-to-noise of the observations only nine of the disks $(17\%)$ have well constrained gas disk sizes. 

Due to uncertainties in stellar evolution models of young stellar objects (YSOs) individual ages of protoplanetary disks are hard to measure. More robust are the age differences between different star-forming regions. In order to compare our models to the observations we therefore assume an age of 1-3 Myr for all disks in Lupus (see, e.g. \cite{comeron2008,alcala2014,alcala2017}) and an age of 5-11 Myr for all disks in Upper Sco (see, e.g. \citealt{Preibisch2002,Pecaut2012}). 

\subsubsection{Wind-driven or viscous disk evolution?}
\label{sec: wind-driven or viscous}

Let us start by comparing representative models of wind-driven (with constant \aDW) and viscous evolution to the observations.
The left panel of Figure \ref{fig: comparing to observations} shows the observed gas disk sizes in Lupus and Upper Sco and compares them to the measured gas disk sizes of our wind-driven model with $\mo=0.01\,\msun$ and $\ro= [20, 65]$ AU. The middle panel does the same for two viscously evolving disk models with $\alpha_{\rm visc} =10^{-3}$ and $10^{-4}$, taken from \cite{Trapman2020}. Note that the initial disk masses of the viscous models were chosen such that the models reproduce the observed stellar accretion rates in Lupus at 1 Myr (see \citealt{alcala2014,alcala2017,Trapman2020}). An estimate of the stellar mass accretion rate of our disk wind model, using $\macc(1\,\mathrm{Myr})\approx\mdisk(1\,\mathrm{Myr})/\tacc$, shows that it has $\macc(1\,\mathrm{Myr})\approx7\times10^{-9}\ \msun\ \mathrm{yr}^{-1}$ which is comparable to the accretion rate used to calculate \mo\ for the viscous models shown here $(\macc = 10^{-8}\ \msun\ \mathrm{yr}^{-1})$.

The left and middle panel of Figure \ref{fig: comparing to observations} highlight the contrast between the two evolutionary theories: \rgas\ of the wind-driven model decreases from $\sim470$ AU at 0.1 Myr to $\sim80$ AU at 10 Myr, whereas the viscous model with $\alpha_{\rm visc} = 10^{-3}$ is almost the direct opposite, increasing from $\sim110$ AU to $\sim560$ AU over a 10 Myr time period. Comparing to the observations, the wind-driven model can explain both the observed disk sizes in Lupus and Upper Sco. Specifically, the model reproduces the fact that disks in Upper Sco are on average smaller than those in Lupus. This last fact in particular distinguishes the wind-driven model from the viscous disk models. Viscous evolution can explain the observed gas disk sizes in Lupus, but it has more difficulty explaining the small disks sizes in Upper Sco, especially while also explaining the larger disks sizes in Lupus. \cite{Trapman2020} suggested that the small disks sizes in Upper Sco could be the result of external photo-evaporation due to the proximity of these disks to the Sco-Cen OB-association. Assuming that disk evolution is driven by MHD disk winds allows us to explain the observations without having to invoke external photo-evaporation.

The comparison between observed gas disk sizes and the representative wind-driven disk models seem to suggest that the evolution of protoplanetary disks is driven by MHD disk winds. There is, however, a caveat that should be mentioned here. The disk wind models that reproduce the observed range in gas disk sizes at 1-3 Myr are much larger at younger ages $(\rgas \approx 100-600\ \mathrm{AU}\ \mathrm{at\ <0.5\ Myr})$. While measurements of gas disk sizes at these young ages are limited, current observations suggest that Keplerian disks larger than 50 AU are rare (see e.g. \citealt{NajitaBergin2018,Maret2020}). Measurements of the dust disk size are more common, and while the dust disk size is not the same as the gas disk size, they can provide some estimate of the size of the disk. In their VANDAM survey, \cite{Tobin2020} presented ALMA 0.87 millimeter continuum observations of Class 0, Class I and Flat Spectrum sources in the Orion star-forming region. From these observations they measured the dust disk size, defined as the deconvolved Gaussian $2\sigma$ radius (equivalent to a radius that encloses 95\% of the flux) from the fits to the continuum images, for 108 Class I embedded disks. They find a mean dust disk size of 35.4$^{+3.5}_{-6.1}$ AU for sources that are not in a multiple system. A similar average dust disk size is found for young sources in the Ophiuchus star-forming region \citep{Cieza2019}.

Properly comparing these observed dust radii to the gas radii of our models is a non-trivial exercise. It would require both radiative transfer calculations as well as the inclusion of the evolution of dust (see, e.g. \citealt{Sheehan2020}). This is beyond the scope of this paper. Instead we limit ourselves to a simple direct comparison where we show how much larger \rgas\ must be than mean Gaussian $2\sigma$ radius of the VANDAM survey $(R_{\rm dust,\ VANDAM})$ to reconcile the disk wind models and the observations of young disks. 

Figure \ref{fig: comparing to observations} shows \rgas\ needs to be $\sim8\times$ larger than the observed mean $R_{\rm dust,\ VANDAM}$ to match the size of the disk wind models at the average age of a Class I disk ($0.16-0.7$ Myr; see e.g. \citealt{Evans2009}).
At first glance this seems at odds with the physical radius $(\ro=65\ \mathrm{AU})$ which is much closer to the mean Class I dust disk size. However, \text{we should} keep in mind that $R_{\rm dust,\ VANDAM}$ encloses most of the continuum emission and lies close to the outer edge of the dust disk. For reference, the radius that encloses 95\% of the mass is $\sim 3\times$ larger than \ro\ for a tapered surface density with a slope of one. Under the simplifying assumption that the continuum emission traces the surface density, we can expect that $R_{\rm dust,\ VANDAM}$ is several times larger than \ro. 
Among Class II disks such a large ratio between gas disk size and dust disk size is a sign of substantial dust evolution, which seems unlikely for these young sources (see, e.g. \citealt{Trapman2019a,Rosotti2019,Toci2021}).  It thus seems difficult to reconcile the picture of wind-driven disk evolution with observed disk sizes at all disk ages.

\subsubsection{Constraining the initial mass and size using observed \rgas}
\label{sec: wind-driven initial mass ans size}

Figure \ref{fig: comparing initial conditions to observations} compares wind-driven model with different initial masses and sizes to the observation in Lupus and Upper Sco. Assuming a fixed initial disk size of $\ro=65$ AU initial disk masses between $10^{-4}\ \msun \lessapprox \mo \lessapprox 10^{-1}\ \msun$ lie with the observed gas disk size range. Along a similar line, if we fix the initial disk mass to $\mo=0.1\ \msun$ models with initial disk sizes between $20\ \mathrm{AU} \lessapprox \ro \lessapprox 90\ \mathrm{AU}$ span the observed range of \rgas. 
We note however that these are only a simple, first order estimate of the initial conditions that would reproduce the observations. For example, not all of these models will reproduce the observed stellar mass accretion rates.

It should be noted here that we compare our models to 56\% of the 1-3 Myr old Lupus disk population and 22\% of the 5-11 Myr old Upper Sco disk population.
To fully answer the question whether disk evolution is driven by disk winds or viscous expansion requires a more complete sample of these regions obtained with deeper observations, such as will be provided by the forthcoming ALMA large program AGEPRO, the inclusion of more star-forming regions of different ages and a population synthesis study that folds in information from other parameters such as stellar mass accretion rate.

\subsubsection{Population synthesis using a $\Sigma_{\rm c}-$dependent \aDW}
\label{sec: population synthesis}

\begin{figure}
    \centering
    \includegraphics[width=\columnwidth]{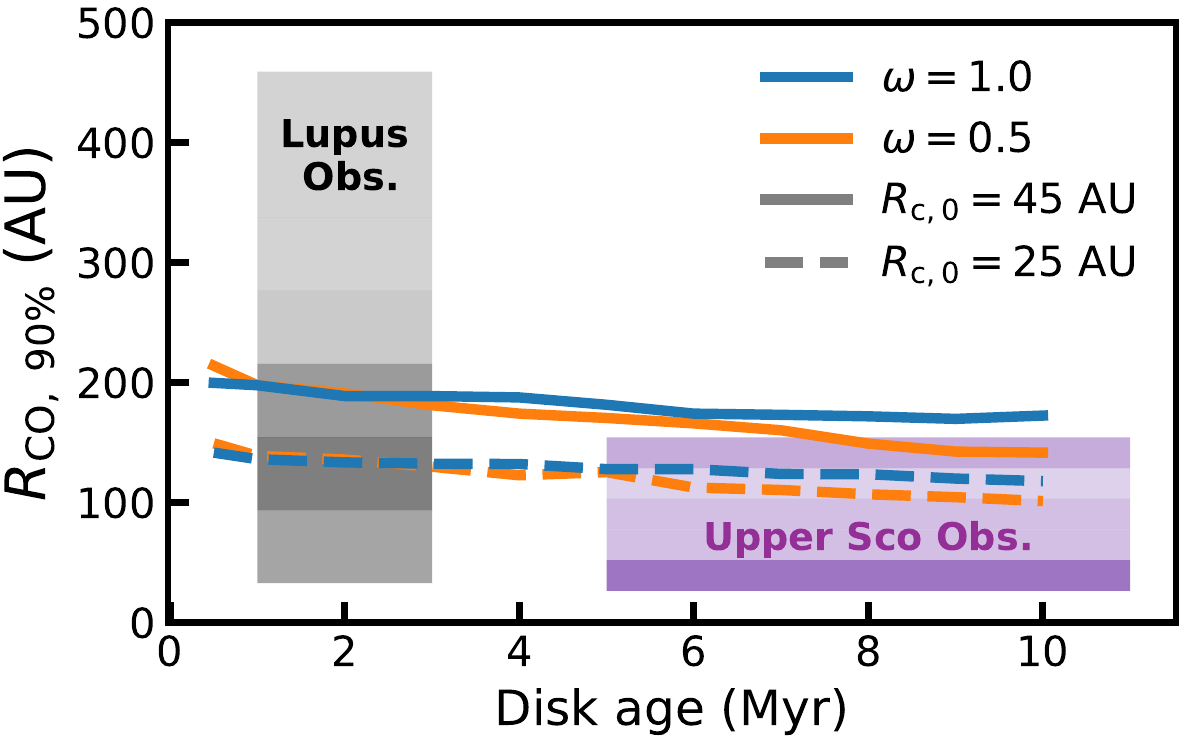}
    \caption{\label{fig: dissipating disks} Evolution of \rgas\ using the evolution of the median disk mass of the disk population synthesis presented in \cite{Tabone2021b} (see their supplemental material). Colors indicate different scalings between \aDW\ and $\Sigma_{\rm c}~(\aDW\propto\Sigma_{\rm c}^{-\omega})$. The solid and dashed lines show initial disk sizes of $\ro = 45$ and 25 AU, respectively. 
    Grey and purple shaded regions show the distribution of observed \rgas\ in Lupus and Upper Sco, respectively (see also Figure \ref{fig: comparing to observations}).}
\end{figure}

Recently, \cite{Tabone2021b} showed, using a disk population synthesis approach, that a wind-driven disk evolution using a $\Sigma_{\rm c}-$dependent \aDW\ can simultaneously explain both the rate of disk dispersal, inferred from the decrease of disk fraction with star-forming region age, as well as the observed correlation between disk mass and stellar mass accretion rate (see, e.g. \citealt{Fedele2010,Manara2016b,Mulders2017,Manara2019}).
Here we examine if their synthetic disk population also reproduces the observed gas disk sizes in Lupus and Upper Sco. We take the synthetic disk population model of \cite{Tabone2021b} to predict the evolution of the median \rgas. This evolution is the result of the evolution of the disk mass and the survivorship bias due to disk dispersal (see their supplemental material).

Figure \ref{fig: dissipating disks} shows \rgas\ versus time overlaid on the observed distribution of \rgas\ of disks in Lupus and Upper Sco. As discussed in \cite{Tabone2021b} the average disk mass decreases slowly over time, which is also reflected in the evolution of \rgas. Over 10 Myr \rgas\ decreases only minimally, from $\sim210$ AU to $\sim140$ AU for models with $\ro=45$ AU and from $\sim150$ AU to $\sim100$ AU for models with $\ro=25$ AU.

Comparing these models to the observations it is clear that they cannot fully explain the difference in size between disks in Lupus and Upper Sco. Based on Figure \ref{fig: dissipating disks} it is conceivable that the $\rgas<200$ AU disks in Lupus could evolve into the disks that are observed in Upper Sco, but that does not explain what happens with the large $\rgas>200$ AU disks in Lupus. 
Explaining both sets of observations with the $\Sigma_{\rm c}$-dependent \aDW\ would require an additional process that reduces the observed disk size over a 5-10 Myr timescale. 

As will be discussed in Section \ref{sec: CO underabundance} a good candidate for such a process is a decrease of the CO abundance over time, caused by either chemical conversion of CO into more complex species or the locking up of CO into larger dust bodies. Based on the bottom panel of Figure \ref{fig: effect CO underabundance - 7mJy cut} a reduced peak CO abundance $\leq 10^{-6}$ would reduce \rgas\ by a factor 4, lowering the \rgas\ presented in Figure \ref{fig: dissipating disks} to similar numbers as seen in observations in Upper Sco. 
Studies show that such a decrease in CO abundance over a 5-10 Myr time span is definitively feasible (see, e.g. \citealt{Yu2016,Yu2017,Bosman2018b,Schwarz2018,Krijt2020,Trapman2021}). In particular, \cite{Anderson2019} showed that CO abundances $\leq10^{-6}$ are required to reproduce the observed N$_2$H$^+$ and CO line fluxes of two disks in Upper Sco.
Furthermore, the disks in Upper Sco are located close to Upper Scorpius OB association. External photoevaporation caused by the abundance of O and B stars near the disks in Upper Sco could have stripped gas from the outer part of the disk, causing these disks to be smaller than expected from just wind-driven disk evolution. 
Deeper observations would allow us to distinguish between these two scenarios. If a low CO abundance is the cause for the small \rgas\ in Upper Sco, deeper observations should reveal the faint CO emission that surrounds these disks and their \rgas\ would increase (see Figure \ref{fig: effect CO underabundance - appendix}). Conversely, external photoevaporation would remove the material that produces the optically thin CO emission and the disks would remain small.

\subsection{Caveats}
\label{sec: caveats}
\begin{figure}
    \centering
    \includegraphics[width=\columnwidth]{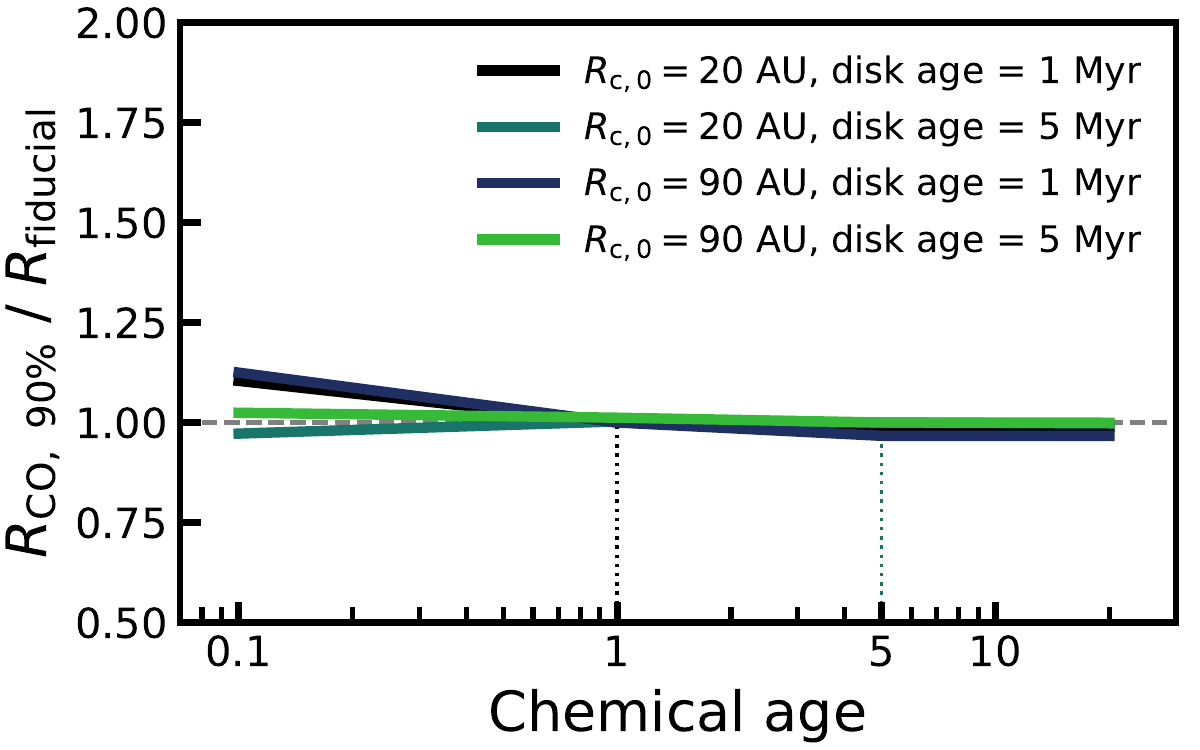}
    \caption{\label{fig: impact of chemistry} Effect of the chemical age on the measured \rgas. Colors show the four models ($\mo=0.1\ \msun, \ro=[20, 90]\ \mathrm{AU}$, disk age = [1,\,5] Myr) that were rerun with a different chemical age. The y-axis shows the ratio of the resulting \rgas\ and the \rgas\ of the original model.}
\end{figure}

In this work we have made several assumptions, such as the values for \tacc\ and $\lambda$ and the shape of the surface density profile (see Section \ref{sec: initial condition}). Here we discuss how these assumptions impact our results.

\subsubsection{The choice of accretion timescale \tacc}
With the exception of the population synthesis in the previous section we have assumed a single accretion timescale $\tacc = 0.5\,\mathrm{Myr}$ for computing the evolution of the surface density in our models (see Section \ref{sec: initial condition}). It is likely that the accretion timescale differs among individual disks, depending on the physical conditions in the disk (e.g. magnetic field strength or temperature structure; see, e.g, \citealt{BaiStone2013,Suzuki2016}, \cite{Tabone2021a}). Note that \tacc\ is constrained observationally by $\mdisk/\macc$ to a few Myr. 

As shown in Equation \eqref{eq: mass evolution}, the accretion timescale only enters in the evolution as the ratio with the age of the disk $t/\tacc$. This suggests that the value of \tacc\ does not matter when evaluating our results against $t/\tacc$. However, the age of the disk also enters in models in the time-dependent chemistry. The chemistry of each model is run up to the age that was used to calculate its surface density (see Section \ref{sec: DALI models}). Here we examine what the effect of the chemical age is on the observed gas disk size \rgas. If the effect is small then sizes measured from our models are valid for any combination of disk age and accretion timescale that matches their current $t/\tacc$.

Figure \ref{fig: impact of chemistry} shows how much changing the chemical age affects \rgas. Four models ($\mo=0.1\ \msun$, $\ro= [20,90]$ AU and disk ages of 1 and 5 Myr) were run again, but now with four different chemical ages: 0.1, 1, 5 and 20 Myr. The newly measured \rgas\ are then compared to the \rgas\ of the original model. The largest differences are seen for short chemical ages, but even then they are within 10\,\%, indicating that the chemical age only has a small impact on the measured \rgas. At longer chemical ages the chemistry converges and \rgas\ no longer changes. 
This is likely because photodissociation is the dominant process in the chemistry of CO in the higher up regions of the disk that are most relevant for the $^{12}$CO 2-1 emission from which we measure \rgas. 

The fact that the chemical age does not significantly affect \rgas\ means that our results can be evaluated against $t/\tacc$. For example, the \rgas\ measured from one of our models with a disk age of 2 Myr ($t/\tacc$ = 2 Myr / 0.5 Myr = 4) will match the \rgas\ of a disk with $\tacc = 2$ Myr at 8 Myr or $\tacc = 0.2$ Myr at 0.8 Myr to within a few percent.
Note that processes such as the chemical conversion of CO over longer timescales are not included in DALI. Their effect will be discussed in more detail in Section \ref{sec: CO underabundance}.

\begin{figure}
    \centering
    \includegraphics[width=\columnwidth]{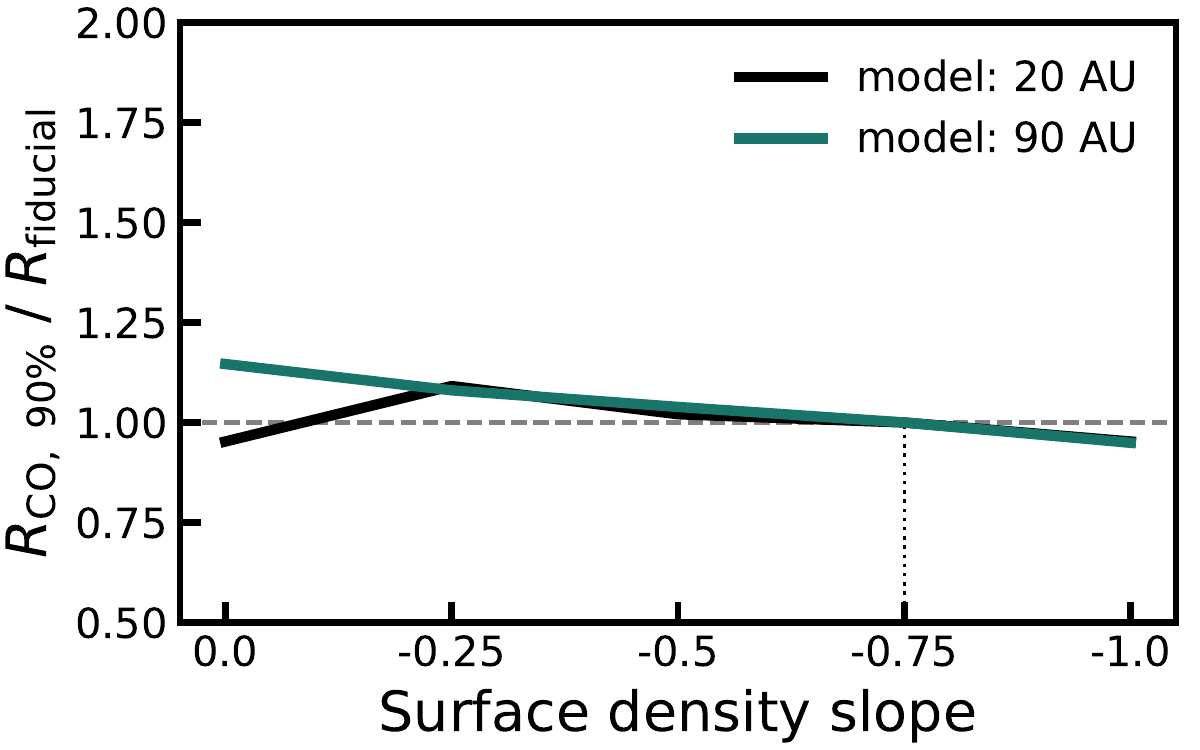}
    \caption{\label{fig: impact of lambda} Effect of the slope of the surface density profile $(\Sigma_{\rm gas} \propto r^x; x=-\gamma + \xi)$ on the measured disk size \rgas. Colors show the two models $(\mo=0.1\,\msun, \ro = [20,90]\,\mathrm{AU},\,\mathrm{disk\ age} = 2\,\mathrm{Myr})$ that were rerun with a different slope. The y-axis shows the ratio of \rgas\ compared to the original model.}
\end{figure}

\subsubsection{The magnetic lever arm $\lambda$}
In a wind-driven disk $\lambda$ regulates the efficiency at which the disk wind is able to extract angular momentum from the disk. In the model, $\lambda$ is one of the parameters that sets the slope of the surface density profile ($\Sigma\propto R^{-\gamma+\xi}; \xi = 1/2(\lambda-1)$, see also Equation \eqref{eq: surface density general}). For our models we have assumed $\lambda=3$ and $\gamma=1$, which results in $\Sigma\propto R^{-0.75}$. As $\lambda \rightarrow \frac{3}{2}$ the slope of $\Sigma$ goes to zero and the surface density becomes constant. As observational constraints of the $\lambda$ parameter are sparse (see, e.g. \citealt{Tabone2017, deValon2020}) 
it is worth investigating how much the slope of the surface density, set by $\gamma$ and $\lambda$, affect the measured \rgas. 

Figure \ref{fig: impact of lambda} shows the \rgas\ of two sets of models $(\mo = 0.1 \msun, \ro = [20, 90]$ AU) that were run with different surface density slopes $(0, -0.25, -0.5, -0.75, -1)$, which corresponds to $\lambda = 1.5, 1.67, 2, 3\ \mathrm{and}\ \infty$, respectively. Despite the large range of slopes covered, the effect on \rgas\ is minimal. Compared to the $\lambda = 3$ models used in the rest of this work \rgas\ changes by less than 15 \%. The main reason for this is that the $^{12}$CO emission from which \rgas\ is measured is optically thick, meaning the emission profile follows the slope of the temperature profile rather than the slope of the surface density.

\subsubsection{Shape of the surface density profile in the outer disk}
As discussed at the end of Section \ref{sec: evolution - cst alpha - single model}, the evolution of \rgas\ is directly linked to the exponential taper of the surface density. For a viscously evolving disk this is a natural feature of the surface density profile. Turbulence in the outer disk will smooth out any initially sharp outer edge into the shape of an exponential taper (see, e.g. \citealt{LyndenBellPringle1974}). However, if the disk evolution is driven by disk winds there is no prior requirement for turbulence in the disk. 
An exponentially tapered outer disk is therefore not guaranteed when disk evolution is driven by disk winds.
Note that a non-zero amount of turbulence in a wind-driven disk would also smooth out the outer disk, even if that turbulence is not the main driver of the evolution. 

Independent of the shape of the surface density in the outer disk the CO intensity will follow the shape of the temperature profile while it remains optically thick. If we assume that the transition from optically thick to optically thin emission lies close to the radius where CO can be photodissociated in the outer disk, which we refer to as $R_{\rm CO,\ p.d.}$, most of the emission will be optically thick and \rgas\ is related to $R_{\rm CO,\ p.d.}$ through 
\begin{equation}
\rgas = 0.9^{(\frac{1}{2-\beta})} R_{\rm CO,\ p.d.},    
\end{equation}
where $\beta$ is the slope of the temperature profile (see Equation 4 in \citealt{Trapman2019a}).
If we further assume that CO becomes photodissociated at a fixed column density the evolution of \rgas\ is directly linked to the evolution of the disk mass through the shape of the surface density in the outer disk.

Under these assumptions we can predict the evolution of \rgas. Let us write the surface density at $R_{\rm CO,\ p.d.}$ as 
\begin{equation}
\Sigma_{\rm CO,\ p.d.}  = \frac{\mdisk}{2\pi\ro^2} f\left(\frac{R}{\ro}\right),
\end{equation}
where $f(x)$ describes the shape of the surface at $R_{\rm CO,\ p.d.}$. For a fixed surface density, this can be rewritten to show that 
\begin{equation}
\rgas \approx R_{\rm CO,\ p.d.} \propto f_{-1} (\mdisk^{-1}),    
\end{equation}
where $f_{-1}(x)$ is the inverse of the function $f(x)$ such that $f_{-1}( f(x)) = x$. For the example of an exponential taper $(f \propto \exp(-x))$ we recover the result from Section \ref{sec: evolution - cst alpha - single model}: 
\begin{align}
\rgas\propto &-\ro \ln \mdisk(t)^{-1}\ro^{-2} \\
             &= \ro\left[ \ln\left(\frac{\mo}{\ro^{2}\Sigma_{\rm CO,\ p.d.}} \right) - \frac{t}{2\tacc}\right].    
\end{align}
If instead the surface density in the outer disk follows a powerlaw with slope $-a\ (f\propto x^{-a})$ we would obtain

\begin{equation}
\rgas\propto \ro^2\mdisk(t)^{\tfrac{1}{a}}.
\end{equation}
Note that when the surface density is changed $\mdisk(t)$ needs to be derived again from Equation \eqref{eq: master equation}. Both changes will directly affect how the evolution of \rgas, meaning
we would need to re-evaluate the conclusions drawn from comparing our models to observations (cf. Section \ref{fig: comparing to observations}). However, this would require more observational constraints on the disk surface density from, for example, spatially resolved CO observations (see, e.g. \citealt{miotello2018}).

It should be noted that we have made several simplifying assumptions here. In particular the assumption that $\rgas\sim R_{\rm CO,\ p.d.}$ is valid if the surface density in the outer disk drops steeply, as is the case for an exponential taper or a sufficiently steep powerlaw. However, if this is not the case optically thin emission in the outer disk could contribute significantly to the total flux and thus affect the location and evolution of \rgas.

\subsection{Impact of a low CO abundance on \rgas}
\label{sec: CO underabundance}

\begin{figure}
    \centering
    \includegraphics[width=\columnwidth]{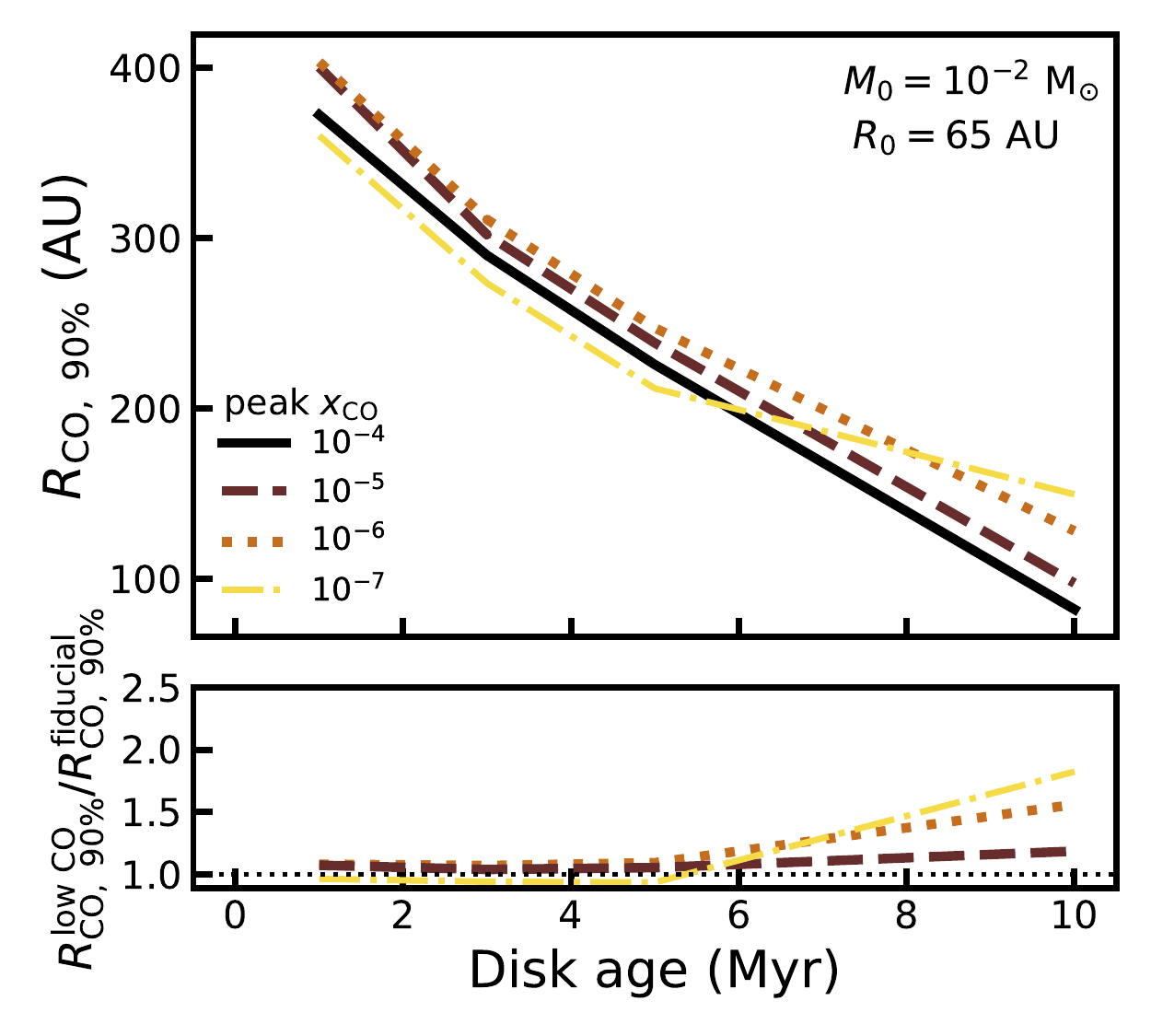}
    \caption{\label{fig: effect CO underabundance - no cut} \textbf{top panel}: \rgas\ as a function of disk age for models with a low CO abundance. Colors and line styles show the peak CO abundance that was put into the model, where a peak CO abundance of $x_{\rm CO} = 10^{-4}$ is our fiducial model. 
    \textbf{Bottom panel}: the ratio of \rgas\ measured from a low CO abundance model to \rgas\ measured from our fiducial model. }
\end{figure}

One of the remarkable findings of recent disk surveys are the low line fluxes and non-detections of optically thin CO isotopologs like $^{13}$CO and C$^{18}$O for most disks. When compared to the dust mass inferred from the continuum luminosity, these faint line fluxes suggest that the majority of disks are either gas poor (compared to the gas-to-dust ratio found in the ISM) or that they are underabundant in CO (but see also \citealt{Miotello2021}, who showed that the faint $^{13}$CO fluxes can be explained by these disks being very compact).
For three disks the gas mass has been measured independently using HD rotational line emission \citep{Bergin2013,McClure2016}. These measurements favor the CO underabundance explanation (see, e.g., \citealt{Favre2013,Kama2016,Schwarz2016,Trapman2017,Calahan2020}), but note that this is a very limited sample, biased towards massive disks. Observed stellar mass accretion rates also argue against low disk gas masses. The low gas masses would imply unreasonably short depletion timescales, in some cases $\leq 10^5$ Myr, suggesting we would be observing all disks just before they disappear (see, e.g. \citealt{Manara2016b,Rosotti2017}).
Two processes have been proposed to explain the underabundance of CO. The first is the chemical conversion of CO into more complex species like for example CO$_2$, CH$_4$ and CH$_3$OH (see, e.g. \citealt{Aikawa1997,Bergin2014,FuruyaAikawa2014,Yu2016,Yu2017,Dodson-Robinson2018,Bosman2018b,Schwarz2018}). Alternatively, CO can freeze out onto dust grains that grow into larger dust bodies that lock up the CO and transport it to smaller radii via radial drift (see, e.g. \citealt{Bergin2010,Bergin2016,Kama2016,Booth2017,Krijt2018}). Note that of course it is likely that both processes contribute (see, e.g., \citealt{Krijt2020}).

Neither of these processes is included in DALI but they do affect the CO from which we measure \rgas, so it is possible, or even likely, that they affect our results. Rather than implementing these processes into DALI (see \citealt{Trapman2021}) we will focus here on their main impact on \rgas, which is reducing the CO abundance. For a set of models $(\ro = 65\ \mathrm{AU}, \mo = 10^{-2}\ \msun)$, we reduce the peak CO abundance $x_{\rm CO}^{\rm peak}$ in the model to $10^{-5}, 10^{-6}\ \mathrm{and}\ 10^{-7}$ (using $x_{\rm CO}^{\rm new} = \min(x_{\rm CO}^{\rm old}, x_{\rm CO}^{\rm peak})$) and recompute the CO excitation and emission. 

We note that removing CO uniformly from our models is a simplification. While there is evidence that efficient vertical mixing smooths out vertical variations (see, e.g. \citealt{Krijt2020,Trapman2021}), both observations and theory show that the underabundance of CO varies radially (see \citealt{Zhang2019,ZhangMAPS2021, Krijt2020}). Given that most of the CO emission is optically thick, the CO abundance in the outer disk, close to \rgas, will have the largest impact on \rgas. The CO abundances discussed here should thus be interpreted as the CO abundance of the outer disk.

\begin{figure}
    \centering
    \includegraphics[width=\columnwidth]{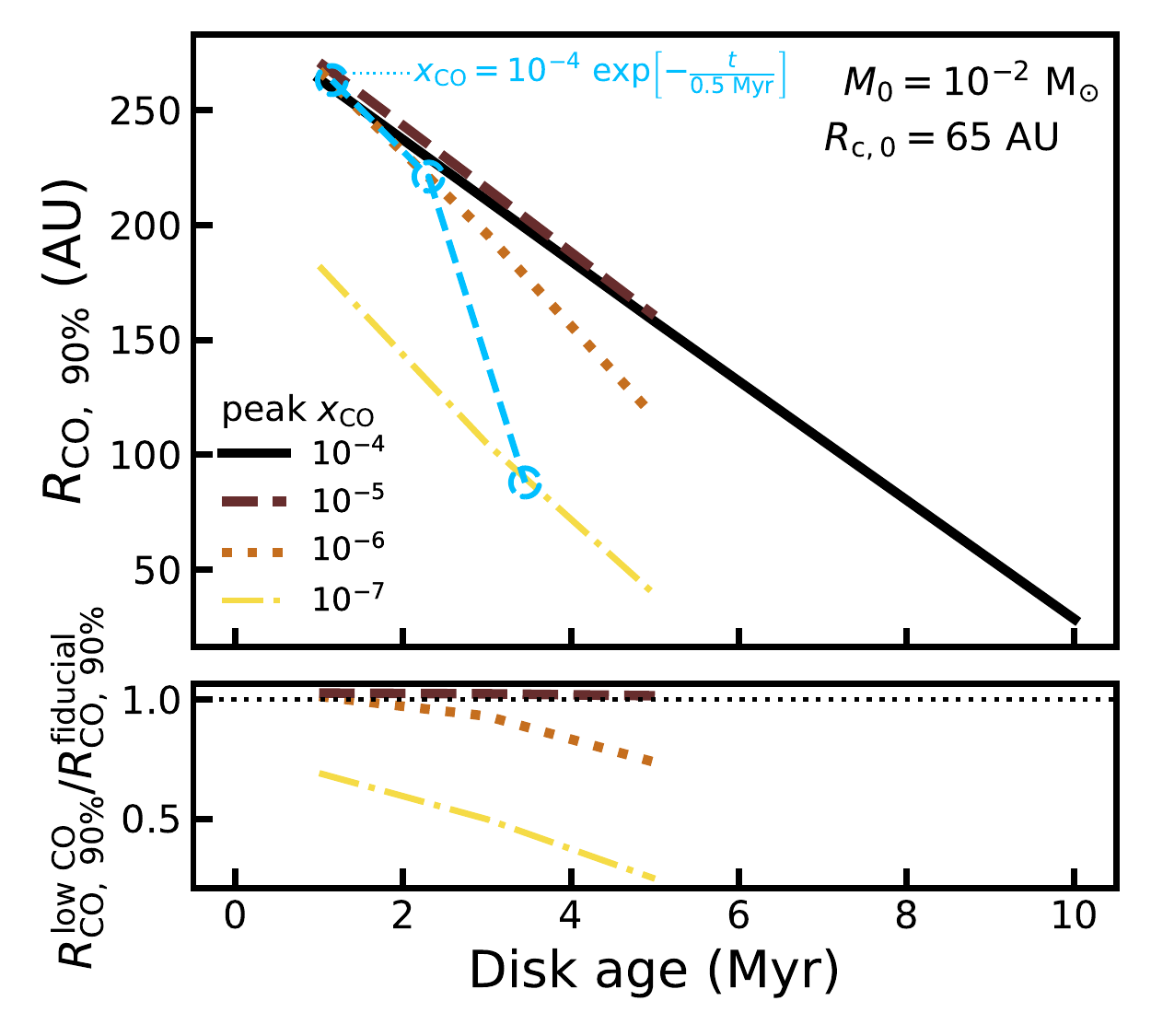}
    \caption{\label{fig: effect CO underabundance - 7mJy cut} as Figure \ref{fig: effect CO underabundance - no cut}, but now accounting for the limited sensitivity of observations. \rgas is calculated after convolving with a 0\farcs37 beam and applying a sensitivity cut of 7 mJy/beam km/s, which approximately mimics the observations of disks in Upper Sco (see \citealt{Barenfeld2016,barenfeld2017}). The top panel shows \rgas\ as a function of disk age. Colors and line styles show the peak CO abundance that was put into the model, where a peak CO abundance of $x_{\rm CO} = 10^{-4}$ is our fiducial model. The blue dashed line shows the track a disk would follow if its CO abundance decreases as $x_{\rm CO} = 10^{-4} \exp(-t/{\rm 0.5\ Myr})$.
    The bottom panel shows the ratio of \rgas\ measured from a low CO abundance model to \rgas\ measured from our fiducial model. Note that for most models no \rgas\ is included at a disk age of 10 Myr, due to the fact that all of the emission from that model lies below our sensitivity cut.}
\end{figure}

Figure \ref{fig: effect CO underabundance - no cut} shows the resulting \rgas\ as a function of disk age.
Up to $\sim5$ Myr models with different CO abundance have a similar \rgas, indicating that decreasing the CO abundance has not significantly affected \rgas. Only after 5 Myr do the models start to diverge. It should be noted that this transition is linked to disk mass, such that for a lower initial disk mass the effects of lowering the CO abundance will be seen at a younger disk age.
Interestingly, the models with a lower $x_{\rm CO}^{\rm peak}$ have a larger \rgas, compared to our fiducial models with $x_{\rm CO}^{\rm peak} = 10^{-4}$. Reducing the CO abundance decreases the CO flux in the inner part of the disk, or more accurately, it shrinks the size of the optically thick emitting area (cf. Section \ref{sec: evolution - cst alpha - single model}), making the emission from the optically thin outer disk by comparison brighter. Given that we have defined \rgas\ as the fraction of the total flux, reducing the flux in the inner disk will move \rgas\ outward (see also \citealt{Trapman2020}). In this way, reducing the CO abundance can increase \rgas\ by a factor of up to 2.5\,. 

However, when comparing to observations we also need to consider the brightness sensitivity of the observations. The emission coming from the outer disk is faint and it is unlikely that all of it would be detected, especially in the 1-3 minute snapshot from which \rgas\ is measured in Lupus and Upper Sco (see \citealt{barenfeld2017,ansdell2018}). To account for the limited surface brightness sensitivity we convolve the moment zero maps of our models with a 0\farcs37 beam and set any emission below a threshold of 7 mJy/beam km/s to zero (see Figure \ref{fig: effect CO underabundance - appendix} in the Appendix for an example). This threshold roughly corresponds to the $1\sigma$ brightness sensitivity of the ALMA observations of disks in Upper Sco (see \citealt{Barenfeld2016,barenfeld2017}). 
From these masked moment zero maps we re-measure \rgas, which are presented in Figure \ref{fig: effect CO underabundance - 7mJy cut}.  Going from a peak CO abundance of $x_{\rm CO}^{\rm peak} = 10^{-4}$ to $10^{-5}$ \rgas\ does not change significantly. Reducing $x_{\rm CO}^{\rm peak}$ further to $\leq 10^{-6}$ shows a much bigger effect, with \rgas\ being up to 70\% smaller compared to the fiducial $x_{\rm CO}^{\rm peak} = 10^{-4}$ models. Even more so, most models have completely disappeared below the threshold at 10 Myr (for reference, $\mdisk(10\,\mathrm{Myr}) = 4.5\times10^{-6}\ \mathrm{and}\ 4.5\times10^{-7}\ \msun$ for the two sets of models shown here).
In Appendix \ref{app: CO underbundance extra figures} we show similar figures using the resolution and sensitivity of the observations in Lupus and both resolutions but with a threshold corresponding approximately 1 hour of integration time (see Figure \ref{fig: effect CO underabundance - appendix}). 

Note that the processes that decrease the CO abundance take time to do so, meaning that we expect $x_{\rm CO}^{\rm peak}$ to decrease over time. This would also affect the evolution of \rgas, increasing the rate at which it decreases over time. To visualize this we have added the trajectory a disk would follow through Figure \ref{fig: effect CO underabundance - 7mJy cut} if its peak CO abundance decreases as $x_{\rm CO}^{\rm peak} = 10^{-4} \exp(-t/{\rm 0.5\ Myr})$. This track shows that \rgas\ would decrease quickly, due to the combined effect of the decreasing disk mass and the lowering of the CO abundance. Note that there are currently few constraints on the timescale over which $x_{\rm CO}^{\rm peak}$ decreases, but observations suggest it occurs quickly (see, e.g. \citealt{Zhang2020a}).  

In conclusion, a low CO abundance ($x_{\rm CO}^{\rm peak} \leq 10^{-6}$) can have a significant effect on \rgas, reducing it by up to 70\% or even making the disk disappear in the noise after $\sim5$ Myr if we take into account the brightness sensitivity of current disk surveys. Note that this also has implications for the comparison between models and observations in Section \ref{sec: comparing to observations}. Constraints on masses and sizes in this section, which do not have a reduced CO abundance, should be considered as lower limits.

\section{Conclusions}
\label{sec: conclusions}

In this work we have combined the analytical model for MHD wind-driven disk evolution presented in \cite{Tabone2021a} with the thermochemical code \texttt{DALI.} We examine how the measured disk size, defined as the radius that encloses 90\% of CO 2-1 total flux \rgas, changes with time in this evolutionary scenario. Our conclusions are summarized below:

\begin{itemize}
    \item As the characteristic size \ro\ of a disk does not evolve in a wind-driven disk, the evolution of \rgas\ is fully set by the evolution of disk mass \mdisk. \rgas\ lines up with the point in the outer disk where the $^{12}$CO emission becomes optically thin. As a result of the decreasing \mdisk, \rgas\ decreases with time for a MHD wind-driven disk as opposed to the increase with time seen for a viscously evolving disk. 
    
    \item We find that \rgas\ decreases linearly with age in our thermo-chemical models if \aDW\ is constant. 
     The slope is set by the initial disk size, where a larger \ro\ results in a steeper decline of \rgas. The initial disk mass \mo\ does not affect the slope. Instead, a larger \mo\ simply shifts the evolution of \rgas\ vertically, resulting in a larger \rgas\ at all times. 
     If \aDW\ is time-dependent $(\aDW(t)\propto\Sigma_{\rm c}(t)^{-1})$ \rgas\ decreases very slowly over most of the disk lifetime, until it rapidly decreases as the disk dissipates.
    
    \item MHD wind-driven evolution with a constant $\aDW$ can explain the observed gas disk sizes of disks in the Lupus and Upper Sco star-forming regions, without having to invoke external photo-evaporation. The \rgas\ of the bulk of the disks in these regions can be reproduced with $\mo \approx 10^{-2} - 10^{-4}\ \msun$ and $\ro\leq20$ AU.
    However, these wind-driven models also require the disks to start out large, which is not corroborated by current observations of young, embedded disks. 
    
    \item For the synthetic disk population presented in \cite{Tabone2021a} that uses a time-dependent \aDW\ and reproduces the observed distribution of \mdisk, \macc\ and disk dispersal, the evolution of the median \rgas\ of the population does not reproduce the observations. Specifically, it overpredicts the median \rgas\ of disks in Upper Sco. A reduced CO abundance, as has been inferred from $^{13}$CO and C$^{18}$O line fluxes for disks in other regions, or external evaporation from nearby massive stars could reconcile the synthetic disk population with the observations.
    
    \item Reducing the CO abundance counter-intuitively increases \rgas. It decreases the bright emission from the inner disk, resulting in the faint outer disk to become a significant part of the total flux, which moves \rgas\ outward. However, if we fold in the limited sensitivity of existing disk surveys, decreasing the CO abundance from $10^{-4}$ to $10^{-6}$ can reduce \rgas\ by up to $70-100\%$.

\end{itemize}

Our work suggest that the observed gas disk sizes of protoplanetary disks are consistent with a MHD wind-driven disk evolution. However, it is unclear how the observed sizes of younger disks fit into this picture. Furthermore, we show that our knowledge of the CO abundance in disks, and how it evolves, directly affects our ability to study disk evolution using observed gas disk sizes, especially at the sensitivity of current surveys. Deeper CO observations of disks in older star-forming regions such as Upper Sco are required to detect the faint outer disk expected for disks with a low CO abundance. Detection of this faint emission is essential for measuring the true extent of gas disk and uncovering what drives disk evolution.


\begin{acknowledgements}
We thank the referee for their constructive comments on this manuscript.
L.T. and K. Z. acknowledges the support of the Office of the Vice Chancellor for Research and Graduate Education at the University of Wisconsin – Madison with funding from the Wisconsin Alumni Research Foundation.
B.T. acknowledges support from the research programme Dutch Astrochemistry Network II with project number 614.001.751, which is (partly) financed by the Dutch Research Council (NWO).
GR acknowledges support from the Netherlands Organisation for Scientific Research (NWO, program number 016.Veni.192.233) and from an STFC Ernest Rutherford Fellowship (grant number ST/T003855/1).
All figures were generated with the \texttt{PYTHON}-based package \texttt{MATPLOTLIB} \citep{Hunter2007}. This research made use of Astropy,\footnote{http://www.astropy.org} a community-developed core Python package for Astronomy \citep{astropy:2013, astropy:2018}.
\end{acknowledgements}

\bibliographystyle{aasjournal}
\bibliography{references}

\begin{appendix}

\section{$^{12}$CO 2-1 intensity profiles}
\label{app: extra figures profiles}

\begin{figure*}[hpt]
    \centering
    \begin{minipage}{0.49\textwidth}
    \includegraphics[width=\textwidth]{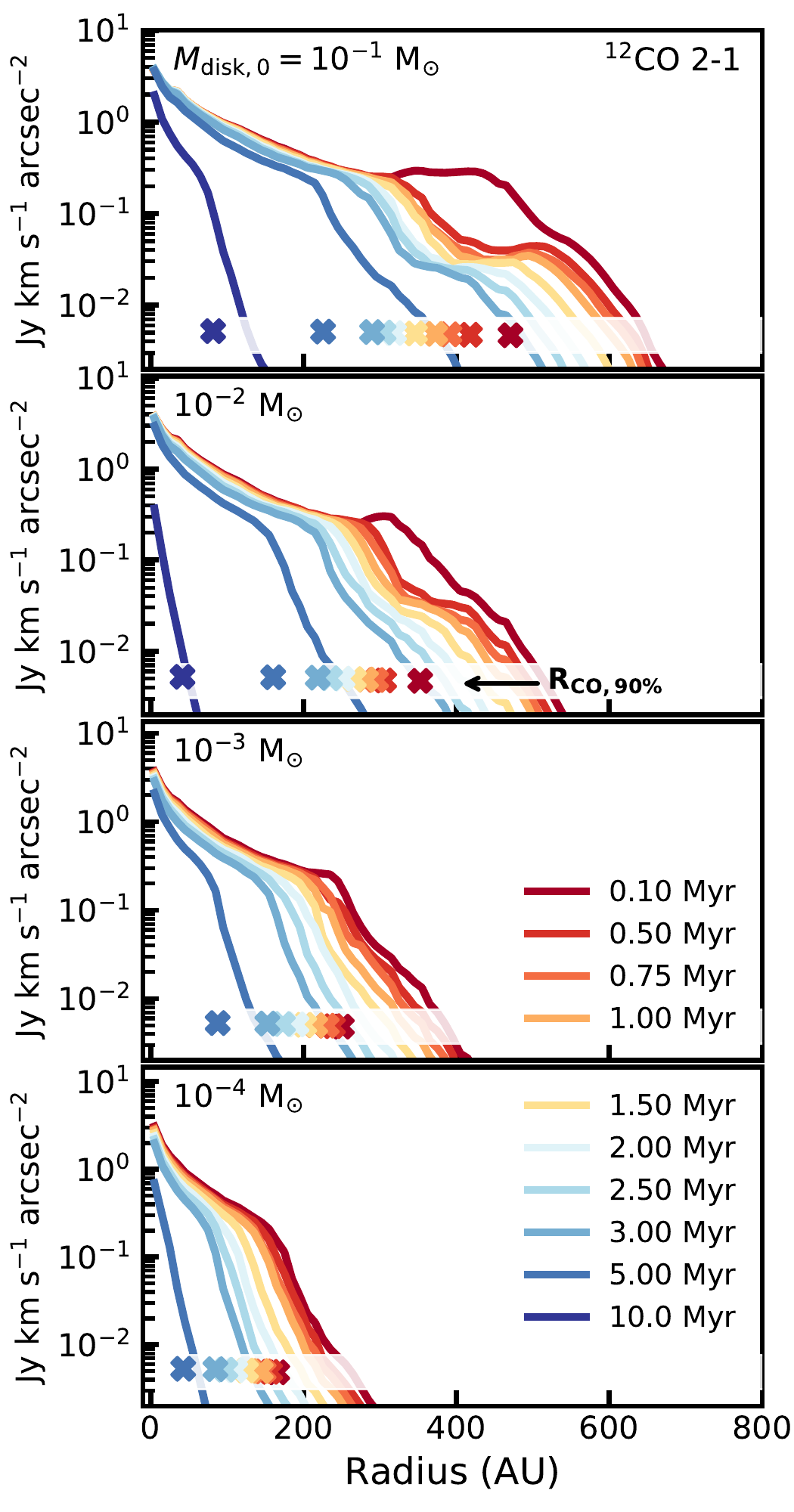}
    \end{minipage}
    \begin{minipage}{0.49\textwidth}
    \includegraphics[width=\textwidth]{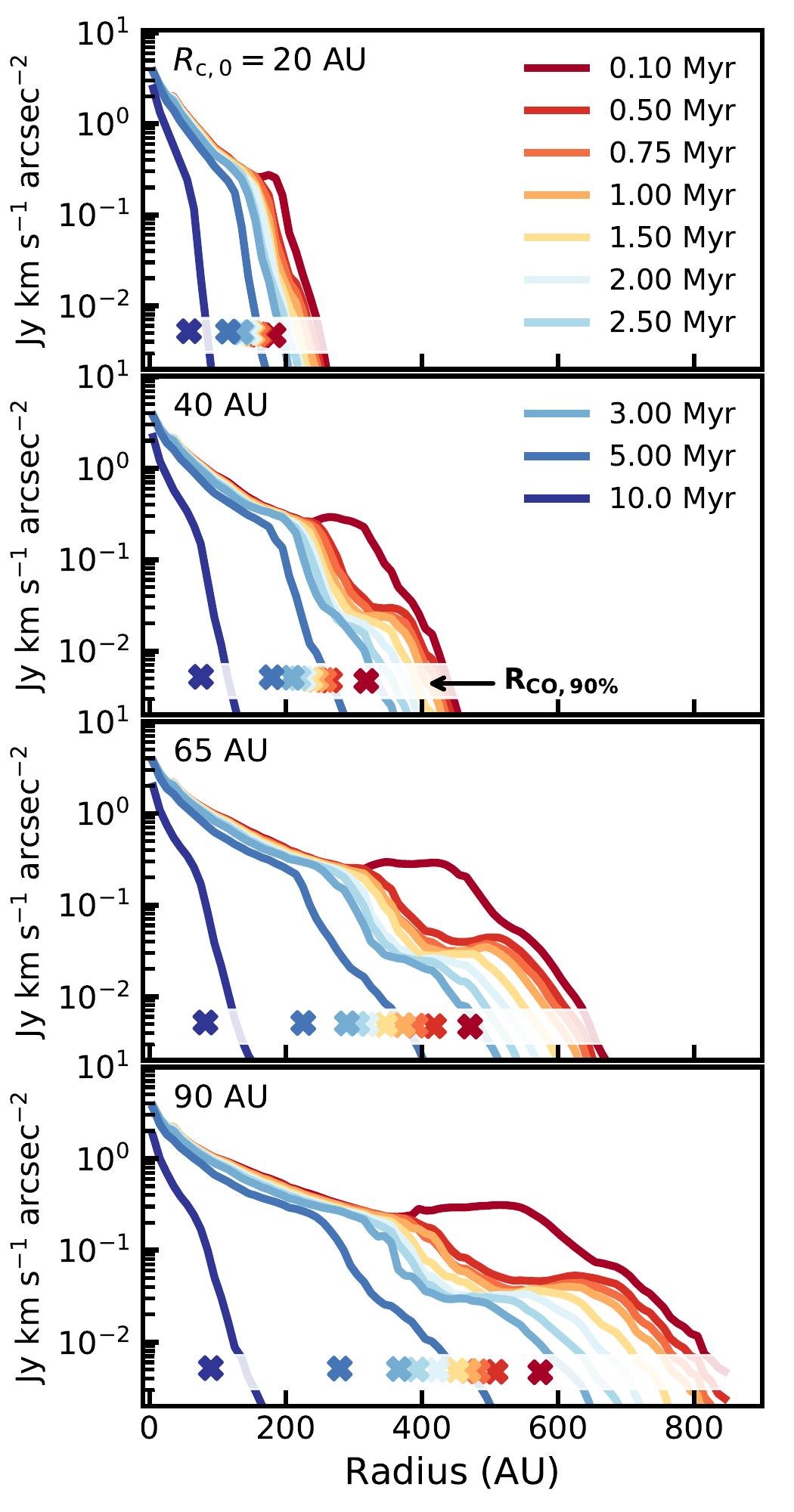}
    \end{minipage}
    \caption{\label{fig: effect of initial conditions} 
    \textbf{Left panels}: Time evolution of the CO 2-1 intensity profile, shown for four wind-driven models with different initial disk masses, decreasing from $\mo=0.1\ \msun$ in the top panel to $\mo=10^{-4}\ \msun$ in the bottom panel. The disk size is kept fixed at $\ro=65$ AU for all models. Colors show different disk ages between 0.1 and 10 Myr. The crosses at the bottom of each panel denote \rgas\ of each model. 
    \textbf{Right panels}: Similar to the left panels, but show models with different initial disk sizes, increasing from $\ro=20$ AU in the top panel to $\ro=90$ in the bottom panel. For these models the initial disk mass is kept fixed at $\mo = 0.1\ \msun$.
    }
\end{figure*}

\section{Extra figures: \rgas/\ro ratio}
\label{app: Extra figure Rgas/Rc}

\begin{figure}
    \centering
    \includegraphics[width=0.5\columnwidth]{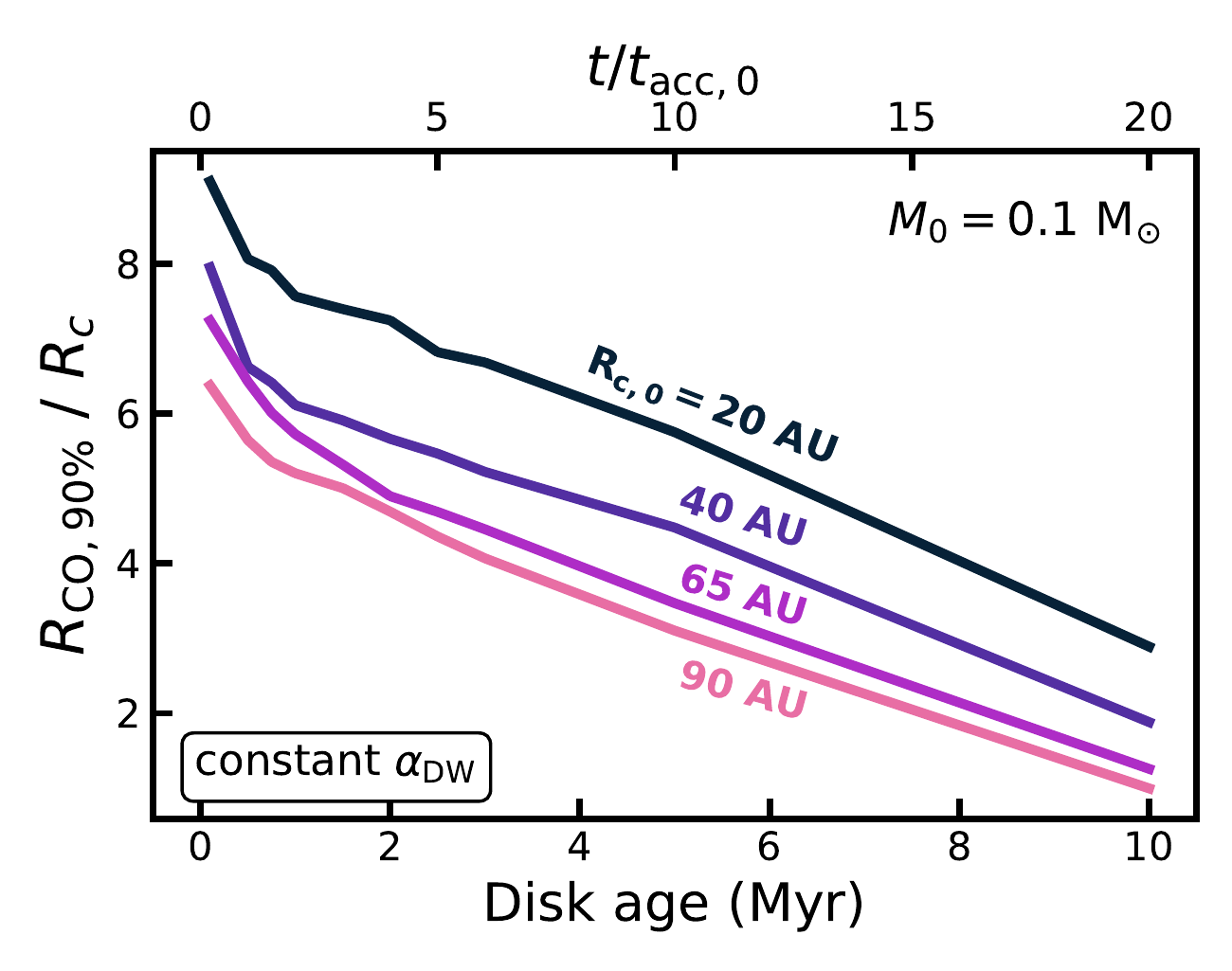}
    \caption{\label{fig: normalized disk size vs time}
    Ratio of \rgas\ over \ro\ versus time for models with different initial different initial disk size (See also the bottom panel of Figure \ref{fig: disk size vs time} in Section \ref{sec: evolution - cst alpha - mass and size}). All models have $\mo=0.1\ \msun$. For reference, the top axis of both panels shows the dimensionless time $t/\tacc$ that goes into the evolution of \mo.   }
\end{figure}

\section{Extra figures: effect of a low CO abundance}
\label{app: CO underbundance extra figures}
\begin{figure}
    \centering
    \begin{minipage}{0.49\columnwidth}
    \includegraphics[width=\columnwidth]{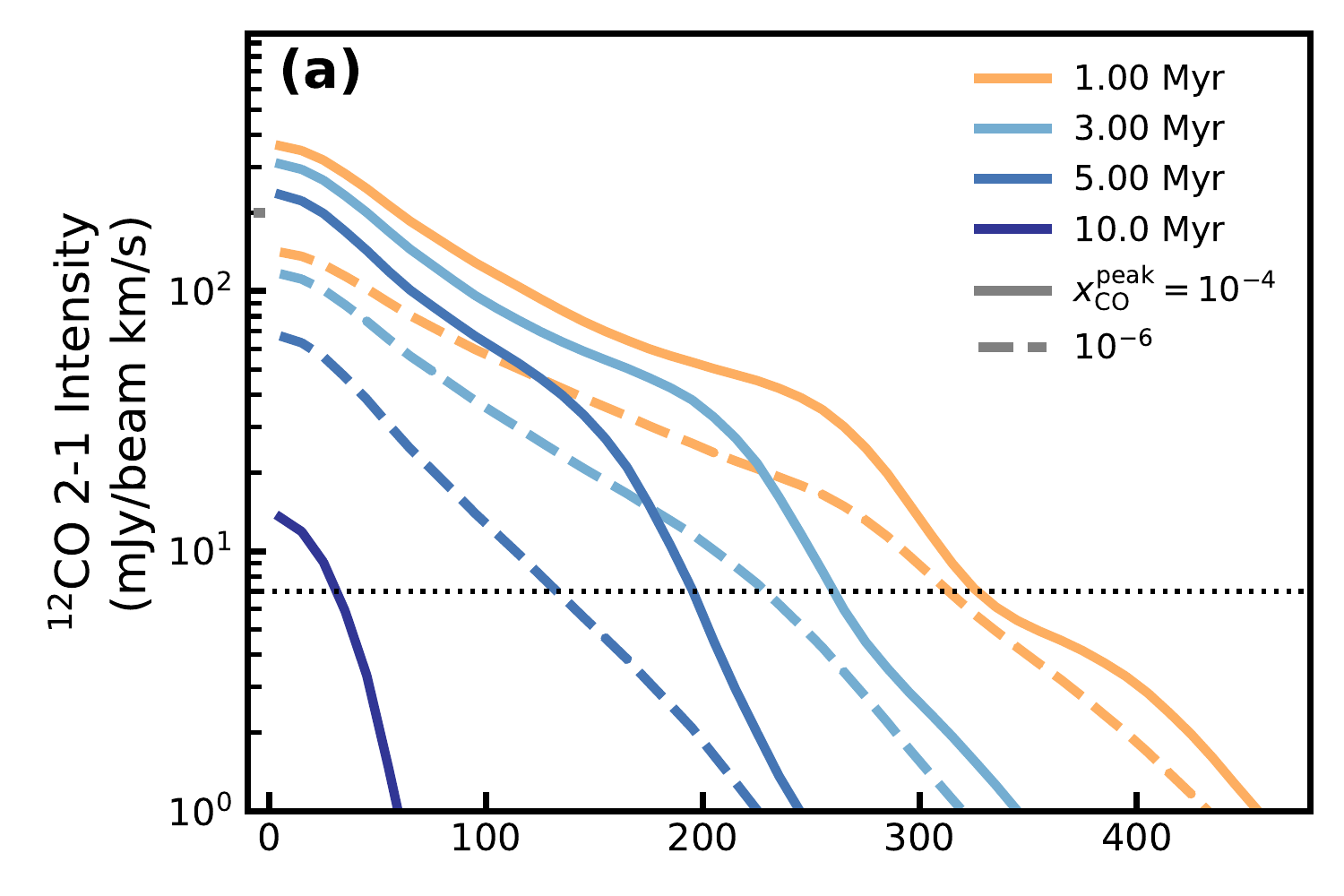}
    \end{minipage}
    \begin{minipage}{0.49\columnwidth}
    \includegraphics[width=\columnwidth]{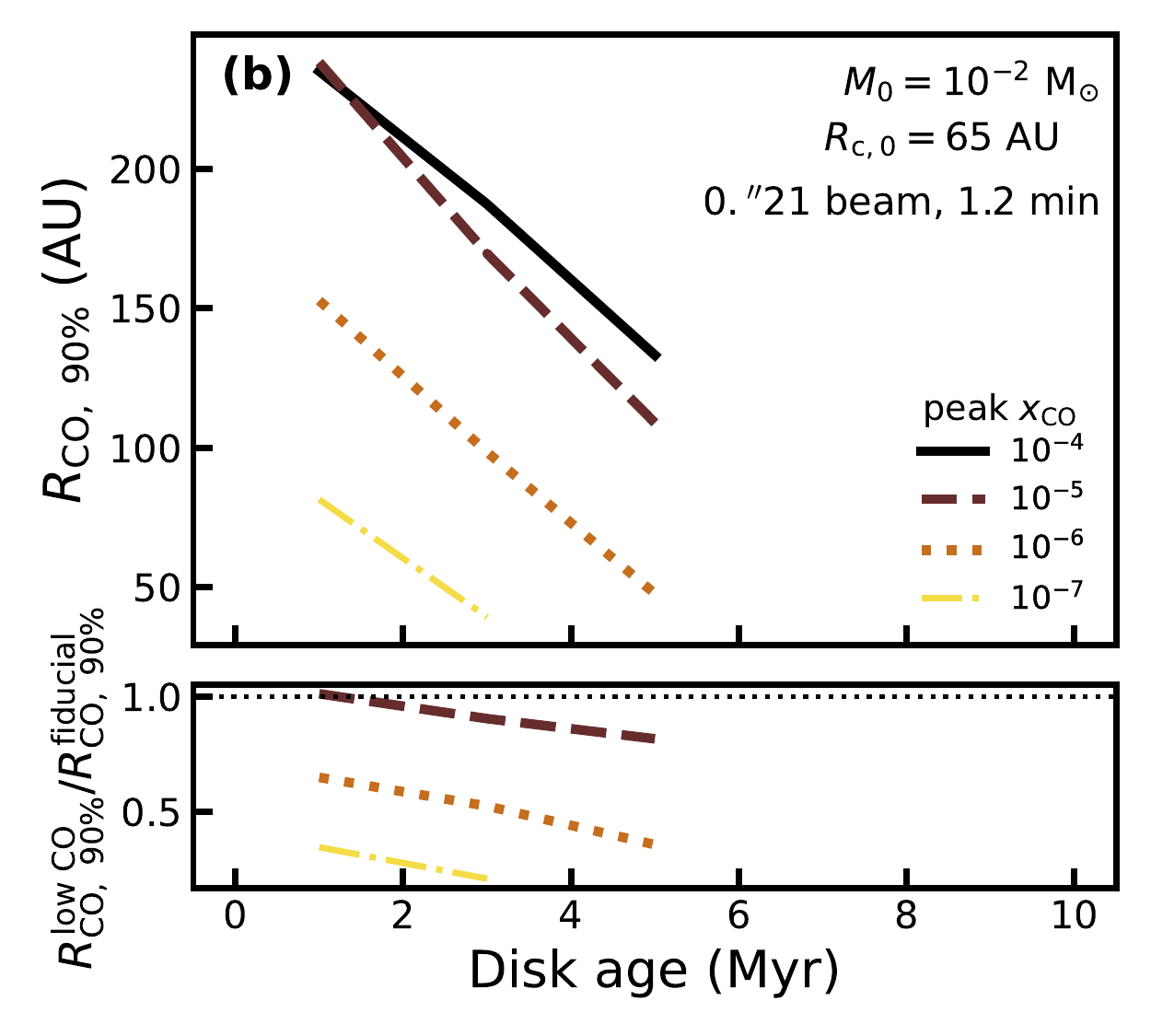}
    \end{minipage}
    \begin{minipage}{0.49\columnwidth}
    \includegraphics[width=\columnwidth]{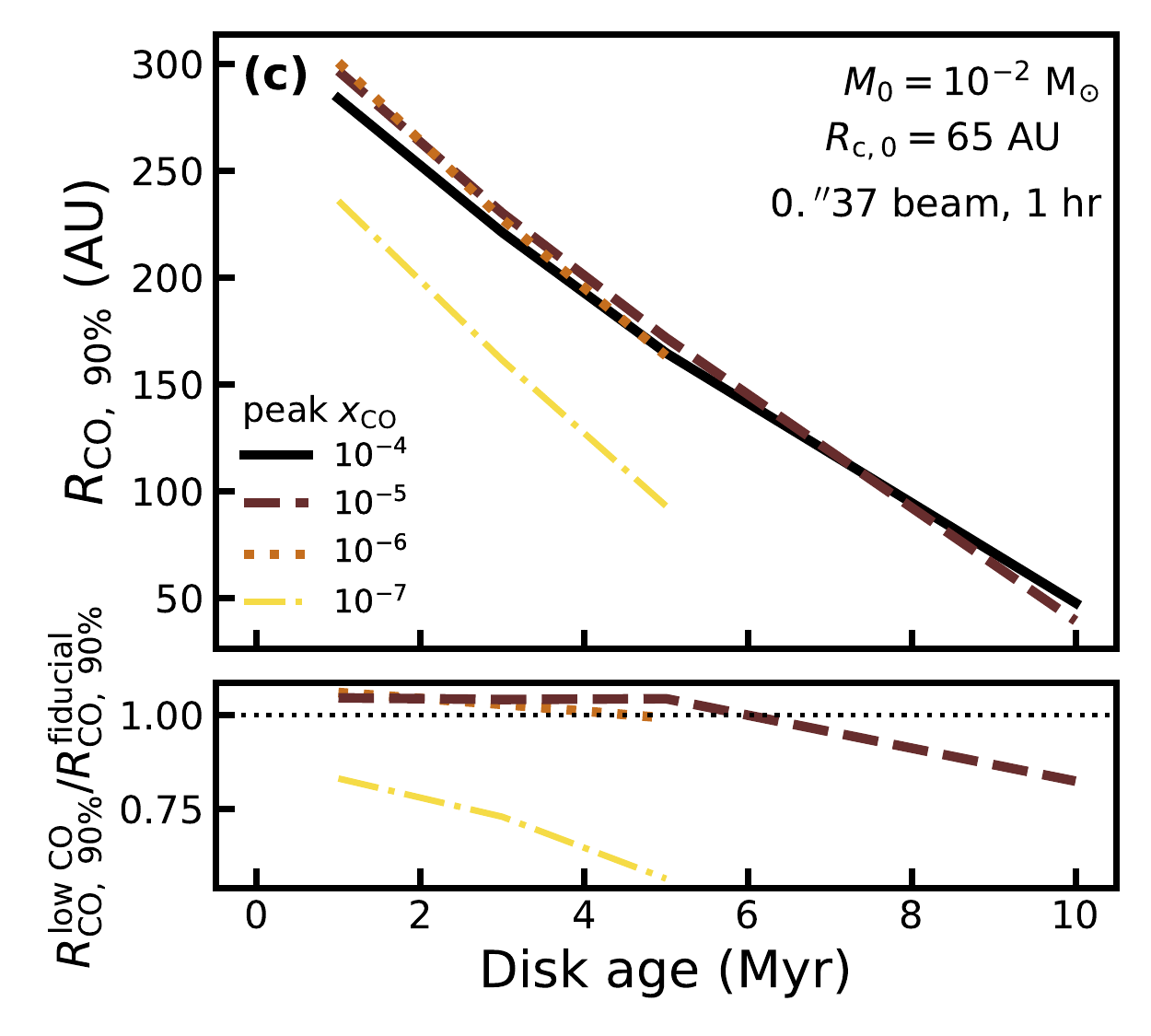}
    \end{minipage}
    \begin{minipage}{0.49\columnwidth}
    \includegraphics[width=\columnwidth]{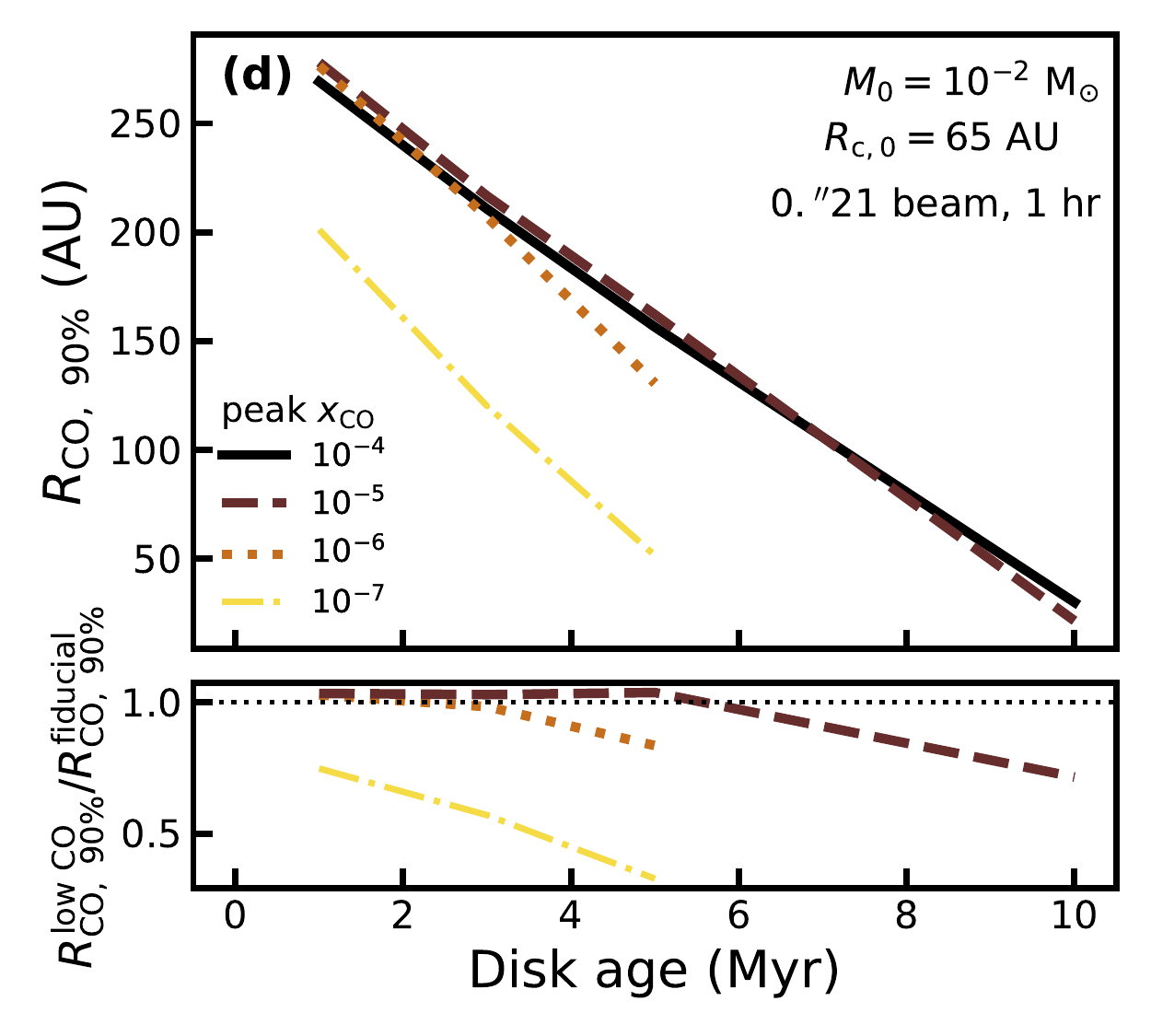}
    \end{minipage}
    \caption{\label{fig: effect CO underabundance - appendix}
    \textbf{Panel (a)}: $^{12}$CO 2-1 intensity profiles for the low CO abundance models presented in Section \ref{sec: CO underabundance}. Solid and dashed lines show models with a peak CO abundance of $10^{-4}$ and $10^{-6}$ respectively. The emission has been convolved with a 0\farcs37 beam. The dotted line shows the brightness sensitivity cut off used to simulate the effect of a limited surface brightness sensitivity in observations.
    \textbf{Panels (b,\,c,\,d)}: \rgas as function of disk age for low CO abundance models. \rgas\ calculated after convolving the synthetic emission map with a beam and applying a sensitivity cut to mimic the limited sensitivity of observations (see Section \ref{sec: CO underabundance}, Figure \ref{fig: effect CO underabundance - 7mJy cut}). Colors and line styles show the peak CO abundance that was put into the model, where a peak CO abundance of $x_{\rm CO} = 10^{-4}$ is our fiducial model.
    Panel \textbf{(b)} matches the beamsize and sensitivity of the Lupus disk survey (see \citealt{ansdell2018}). Panels \textbf{(c)} and \textbf{(d)}, respectively, show. the effect of increasing the integration time of the Upper Sco and Lupus disk survey to 1 hour.
    }
\end{figure}

\end{appendix}

\end{document}